\documentclass[aps,prd,preprint,superscriptaddress,nofootinbib]{revtex4-1}
\usepackage{chay,graphicx}

\newcommand{\n}{\overline{n}}

\newcommand{\eps}{\epsilon}
\newcommand{\ovl}{\overline} 
\newcommand{\mP}{\mathcal{P}}
\newcommand{\chin}{\chi_n}

\newcommand{\euv}{\epsilon_{\mathrm{UV}}}
\newcommand{\eir}{\epsilon_{\mathrm{IR}}}

\newcommand{\blpu}[1]{{\bf{#1}}^{\perp}}

\newcommand{\mc}{\mathcal}
\newcommand{\mr}{\mathrm}

\newcommand{\be}{\begin{equation}} 
\newcommand{\ee}{\end{equation}} 
\newcommand{\bea}{\begin{eqnarray}} 
\newcommand{\eea}{\end{eqnarray}} 

\begin{document}

\title{Factorization of the dijet cross section in electron-positron annihilation with  jet algorithms}

\def\Seoultech{Institute of Convergence Fundamental Studies and School of Liberal Arts, Seoul National University of Science and 
Technology, Seoul 139-743, Korea}
\author{Junegone Chay}
\email[E-mail:]{chay@korea.ac.kr}
\affiliation{Department of Physics, Korea University, Seoul 136-713, Korea}
\author{Chul Kim}
\email[E-mail:]{chul@seoultech.ac.kr}
\affiliation{\Seoultech\vspace{0.5cm}} 
\author{Inchol Kim}
\email[E-mail:]{vorfeed@korea.ac.kr}
\affiliation{Department of Physics, Korea University, Seoul 136-713, Korea}

\begin{abstract} \vspace{0.1cm}\baselineskip 3.0 ex 
 We analyze the effects of jet algorithms on each factorized part of the dijet cross sections in $e^+ e^-$ annihilation using
the soft-collinear effective theory. The jet function and the soft function with a cone-type jet algorithm and the Sterman-Weinberg
jet algorithm are computed to next-to-leading order in $\alpha_s$, and are shown to 
be infrared finite using pure dimensional regularization.  
The integrated and unintegrated jet functions are presented, and compared with other types of jet functions. 
 
\end{abstract}

\maketitle

\baselineskip 3.0 ex 

\section{Introduction}
 
The description of high-energy scattering processes depends on what is measured in experiments.
In the inclusive scattering in which   physical observables  do not depend on the identity of 
the final-state particles, we sum over all the possible configurations of the particles at the parton level, or at 
the hadron level. If we consider the exclusive scattering in which the observables depend on the final-state
hadrons, the hadronization process through the fragmentation functions should be implemented.  
In all the processes, factorization theorems constitute an important  ingredient to separate  nonperturbative effects from hard scattering. 

The factorization proof of high-energy scattering processes has been of great theoretical interest, and the factorization in full QCD was considered
in the pioneering work of Collins, Soper and Sterman (see Refs.~\cite{Collins:1987pm,Collins:1989gx} and 
references therein). Recently the
advent of the soft-collinear effective theory (SCET)  \cite{Bauer:2000ew,Bauer:2000yr,Bauer:2001yt} 
facilitates the proof   in various scattering processes with the systematic treatment of power corrections. 
For example, the factorized dijet cross section in $e^+ e^-$ 
annihilation can be schematically given by the convolution of the hard, collinear and  soft parts as
\begin{equation}
\sigma = H(Q^2, \mu)   J_n (\mu) \otimes J_{\overline{n}} (\mu)  \otimes S (\mu),
\end{equation}
where $H$ is the hard coefficient, $J_n$ and $J_{\overline{n}}$ are the collinear jet functions which describe the final-state energetic collinear 
particles in the lightlike directions $n$ and  $\overline{n}$. The soft function $S$ describes the soft interaction.  In $pp$ scattering, 
the scattering cross section is extended to be given as
\begin{equation}
\sigma = H(Q^2, \mu)   J_n  (\mu) \otimes J_{\overline{n}} (\mu)\otimes f_p (x_1,\mu)\otimes f_p (x_2,\mu) \otimes S (\mu),
\end{equation}
where the additional $f_p$ comes from the parton distribution functions for initial hadrons.

Once the factorization theorem is established, each part can be computed perturbatively and can be resummed by solving the renormalization 
group equation.  However, there are many constraints for the theoretical computation to be consistent in order to satisfy the factorization theorem. 
First,  each part should be infrared safe. Otherwise, the factorization itself does not have any physical meaning. Secondly, since 
the scattering cross section is independent of an arbitrary renormalization scale, the sum of the anomalous dimensions of all the factorized parts should
cancel at a given order in perturbation theory. 

In high-energy processes, final-state hadrons usually form a collimated beam which is called a jet. 
Jets provide 
important information on the physics within and beyond Standard Model. In describing high-energy collisions in terms of jets,  the factorization property is also essential to theoretical predictions. 
There are  diverse ways to define jets depending on the kinematics of the scattering, the detector design, etc.. The jet definitions are represented
by jet algorithms such as the Sterman-Weinberg (SW) algorithm \cite{Sterman:1977wj}, the cone algorithm \cite{Salam:2007xv}, 
the JADE algorithm \cite{Bethke:1988zc}, the $k_T$ algorithm \cite{Catani:1991hj}, the anti-$k_T$ algorithm\cite{Cacciari:2008gp},  to name a few. 

The first theoretical attempt to implement the jet algorithm at the parton level is the Sterman-Weinberg jet. The issue is how to combine adjacent particles into a jet without ambiguity. The important ingredients for a good jet algorithm are that 
it can be employed conveniently both in experiment and in theory. And it should 
be infrared safe. However, implementing a convenient jet algorithm in experiment is sometimes difficult to realize in theory, and vice versa. 
 
In this paper, we first establish the factorization theorem in the dijet production from electron-positron 
annihilation in the framework of SCET.  The cone-type algorithm suggested by Ellis et al.~\cite{Ellis:2010rwa} and the Sterman-Weinberg 
algorithm are employed as specific examples. 
We explicitly show that each factorized part is IR safe at order $\alpha_s$. This is verified in two ways. First 
we compute each factorized part,  using the dimensional regularization both for the ultraviolet (UV) 
and the infrared (IR) divergences.  And we can put a rapidity regulator in the collinear Wilson line to regulate
the IR divergence, while the UV divergence is still handled with the dimensional regularization.

We consider two kinds of jet functions. One is the unintegrated jet function, and the other is the integrated jet function 
which is obtained by integrating the unintegrated jet function with respect to the invariant mass squared of the collinear part. 
Intuitively, the integrated jet function
should be obtained by integrating the unintegrated jet function, which is true in our calculation. Related to this topic, different kinds of 
jet functions have been discussed in Ref.~\cite{Ellis:2010rwa}. The measured jet function describes the jet with thrust, while the unmeasured 
jet function describes the jet without measuring the thrust. However, the 
unmeasured jet function is not obtained by simply integrating the measured jet function. This may look confusing at first sight, 
but this results from different ways of power counting in the measured and unmeasured jet functions. And in deriving physical results such as
the scattering cross section, using the jet functions in both methods are consistent. This 
will be explained in detail.

The idea of applying SCET to the factorization of the dijet cross section has been tested in various directions. An interesting attempt was initiated by Cheung et al.~\cite{Cheung:2009sg} where the authors carefully analyzed 
the phase space according to jet algorithms. The SW, JADE, and $k_T$ jet algorithms were
analyzed  and the dijet cross sections were computed.
  Though they did not explicitly state the factorization theorem, they 
implicitly worked in the framework of SCET by computing the collinear and the soft parts.

Jouttenus~\cite{Jouttenus:2009ns} focused on the  jet functions in $e^+e^-$ annihilation with the factorization scheme in SCET with the SW jet algorithm. 
In Ref.~\cite{Trott:2006bk}, the jet cross section was computed by deriving jet operators and computing the matrix elements of those operators.
In Ref.~\cite{Ellis:2010rwa}, jet shapes were comprehensively analyzed employing SCET.
They used a cone-type jet algorithm and a $k_T$ algorithm as examples to explore the properties of the ``unmeasured'' 
(corresponding to ``integrated"  in our terminology) and the measured (unintegrated) jet functions, as well as the soft functions.  
They pointed out that the unmeasured jet function is not obtained 
by integrating the measured jet function because a different power counting was involved.  This will be 
discussed in detail compared to the fact that our integrated jet function is obtained by integrating the 
unintegrated jet functions.

The structure of the paper is the following: In Section 2, a factorization theorem for the dijet cross section in $e^+ e^-$ annihilation is presented. 
In Section 3, the characteristics of the generic cone-type and the SW jet algorithms are described, and the available phase space for the collinear and soft parts is clarified. We also delineate the power counting of 
the various 
quantities appearing in the factorization theorem. In Section 4, the integrated jet functions with both jet algorithms are computed using the 
dimensional regularization both for the UV and IR divergences, and the unintegrated jet functions are presented in Section 5. In Section 6, the soft function is computed  using the dimensional regularization to
order $\alpha_s$ in both jet algorithms. In Section 7, we verify the SW dijet cross section in SCET is the same as that in full QCD 
at order $\alpha_s$, and  show that the anomalous dimensions of the 
factorized parts are summed to cancel. In Section 8, we discuss the relation between our integrated, unintegrated jet functions and the 
unmeasured, measured jet functions and explain why both approaches are consistent. Finally in Section 9,
we give a conclusions and
perspectives. In Appendix, the jet function in the SW jet algorithm is computed using the rapidity regulator, 
in which the UV divergence is handled by the dimensional 
regularization, while the rapidity divergence is handled by the rapidity regulator.

\section{Factorization of the dijet cross section in SCET}
In $e^+ e^-$ annihilation, as well as in hadron-hadron scattering, a great number of jets may be produced at high energy. A jet algorithm should be
implemented to identify a scattering event with $N$ jets. A proper jet algorithm should separate each collimated beam of particles, and
merge nearby jets without ambiguity with infrared safety. However, in this paper, we are interested in the configuration in which there are 
two back-to-back jets with soft particles, which is called a dijet event. The property of $N$-jet events, for example, $N$ jettiness on event shape, 
can be found elsewhere \cite{Ellis:2010rwa,Stewart:2010tn}.

We consider the dijet cross section in $e^+ e^- \rightarrow j_1 j_2 X$, where $j_1$ and $j_2$  
describe collimated beams of particles.  Because $j_1$ and $j_2$ are  almost back-to-back in the center-of-mass (CM) frame, we 
 describe collinear particles inside $j_1$ and $j_2$ in terms of the lightcone vectors $n$ and $\overline{n}$ respectively 
with $n^2=\n^2=0,~n\cdot\n=2$. 
The dijet scattering cross section is given by
\begin{equation}
\sigma (e^+ e^- \rightarrow j_1 j_2 X) = -\frac{2\pi \sigma_0}{N_c Q^2} \sum_X (2\pi)^4 \delta^{(4)} (q-p_{X_n} -p_{X_{\overline{n}}} 
-p_{X_s} ) \langle 0|J_{\mu}^{\dagger} |X\rangle \langle X|J^{\mu} |0\rangle,
\end{equation}
where $Q^2 = q^2$ is the center-of-mass squared of the $e^+ e^-$ pair. 
Here $\sigma_0$ is the Born cross section for a given flavor $f$ of the quark-antiquark pair, given by
\begin{equation}
\sigma_0 = \frac{4\pi \alpha^2 Q_f^2 N_c}{3Q^2}.
\end{equation}
The final states denoted by $X$ consist of two groups of collinear particles in the $n$ and $\overline{n}$ directions respectively,
 and the remaining soft particles. 
\begin{equation} \label{decomp}
|X\rangle = |X_n\rangle \otimes |X_{\overline{n}}\rangle \otimes |X_s\rangle.
\end{equation}
The factorization theorem can be proved in terms of the hadronic state $|X\rangle$ without this decomposition \cite{Bauer:2008dt}, but 
we use Eq.~(\ref{decomp}) for simplicity.
The collinear and soft particles will be combined into jets according to the jet algorithms, which are 
described in the next section.

In SCET, the collinear particles in the  $n$ and $\overline{n}$ directions have the momentum scaling as
\begin{eqnarray}
p_n^{\mu} &=& (\overline{n}\cdot p_n, p_n^{\perp}, n\cdot p_n ) = (p_n^-, p_{n}^{\perp}, p_n^+) \sim Q(1,\lambda, \lambda^2),
 \nonumber  \\
\label{powc} 
p_{\overline{n}}^{\mu} &=&   (p_{\overline{n}}^-, p_{\n}^{\perp}, p_{\overline{n}}^+) \sim Q(\lambda^2,\lambda, 1),
\end{eqnarray}
where $\lambda$ is a small parameter in SCET, while the soft momentum scales as
\begin{equation}
p_s^{\mu} =(p_s^-, p_{s}^{\perp}, p_s^+) \sim Q(\lambda^2, \lambda^2, \lambda^2).
\end{equation}
For the photon exchange, the electromagnetic current $J^{\mu}$ is written at leading order in SCET as
\begin{equation}
\label{current}
J^{\mu} = C(Q^2,\mu) \overline{\chi}_n \tilde{Y}_n^{\dagger} \gamma^{\mu} \tilde{Y}_{\overline{n}} \chi_{\overline{n}},
\end{equation}
where $\chi_n$ is a gauge-invariant collinear quark field combined with a collinear Wilson line, $\chi_n = W_n^{\dagger} \xi_n$. 
$\tilde{Y}_n (x)$ is the soft Wilson line defined as \cite{Chay:2004zn} 
\begin{equation}
\label{tsoft} 
\tilde{Y}_n (x) = \mathrm{P} \exp \Bigl[ig\int^{\infty}_x ds n\cdot A_s (sn)\Bigr]_, 
\end{equation}  
where `P' denotes the path ordering. Note that the specified path in $\tilde{Y}_{n}$ is different from the standard case of
 $Y_n(x)$~\cite{Bauer:2001yt}, where the path is given by $[-\infty,x]$ along the lightcone vector $n$. 
In Eq.~(\ref{current}), $C(Q^2,\mu)$ is the Wilson coefficient obtained from the matching between full QCD and SCET. The hard coefficient in the scattering 
cross section is given by $H(Q^2,\mu) =|C(Q^2,\mu)|^2$, which is given to one loop as \cite{Manohar:2003vb}
\begin{equation} \label{hard}
H(Q^2,\mu) = 1+\frac{\alpha_s C_F}{2\pi}
\Bigl (-\ln^2  \frac{\mu^2}{Q^2} -3 \ln \frac{\mu^2}{Q^2} -8
+\frac{7\pi^2}{6}\Bigr). 
\end{equation}

We define the (unintegrated)  jet function $J_n$ in the $n$ direction as 
\begin{equation}
\sum_{X_n} \langle 0| \chin^{\alpha} |X_n\rangle \langle X_n | \ovl{\chi}_{n}^{\beta}|0\rangle =
\int \frac{d^4 p_{X_n}}{(2\pi)^3} \n\cdot p_{X_n}\frac{\FMslash{n}}{2}  J_n (p_{X_n}^2,\mu) \delta^{\alpha\beta},
\end{equation}
while the jet function $J_{\overline{n}}$ in the $\overline{n}$ direction is obtained by switching $n$ and $\overline{n}$. 
The jet function at lowest order  is normalized as $J_n^{(0)} (p^2) =  \delta (p^2)$. 
Because the collinear fields do not interact with the soft fields any more after the field redefinition \cite{Bauer:2001yt}, 
the scattering cross section can be written as
\begin{eqnarray}
\sigma &=& \sigma_0  H(Q^2, \mu)\frac{2}{\pi} \int  d^4 p_1 d^4 p_2 \frac{p_1^- p_2^+}{Q^2} 
 J_n (p_1^2,\mu) J_{\overline{n}} (p_2^2,\mu)\nonumber \\
\label{xsec1}
&\times&\sum_{X_s} \delta^{(4)} (q-p_1 -p_2 -p_s) 
\frac{1}{N_c} \mathrm{Tr} \langle 0|\tilde{Y}_{\overline{n}}^{\dagger} \tilde{Y}_n |X_s\rangle \langle X_s| \tilde{Y}_n^{\dagger} \tilde{Y}_{\overline{n}} |0\rangle, 
\end{eqnarray}
where we put $p_1 = p_{X_n}$, $p_2 = p_{X_{\n}}$, and $p_s = p_{X_s}$.
We choose the coordinate system such that $p_1^{\perp}= 0$, then the phase space integral for $p_1$ can be written as
\begin{equation}
\int d^4 p_1 = 2\pi \int dp_1^- dp_1^+ \Bigl(\frac{p_1^--p_1^+}{2}\Bigr)^2 \sim \frac{\pi}{2} \int dp_1^- dp_1^+ (p_1^-)^2.
\end{equation}

In the CM frame,  the delta function in Eq.~(\ref{xsec1}) is rewritten as 
\begin{equation}\label{delta} 
\delta(q-p_1-p_2-p_s) = 2 \delta(Q-p_1^--p_2^--p_s^-)\delta(Q-p_1^+-p_2^+-p_s^+)
\delta^{(2)}(\blpu{p}_1+\blpu{p}_2+\blpu{p}_s).
\end{equation}
Applying the power counting in Eq.~(\ref{powc}) we can simplify further the delta function in Eq.~(\ref{delta}). 
 If we consider a differential scattering cross section for observables of order $\lambda^2$, the soft part is correlated with the collinear parts. 
For example, if we  look at some distribution concerning the range of $\mc{O}(\lambda^2)$ in the $p^-$ component, 
we have to keep $p_{2}^- \sim \mathcal{O} (\lambda^2)$ in $\delta(Q-p_1^--p_2^--p_s^-)$ when $Q-p_1^- \sim \mc{O}(\lambda^2)$. 

Unless we consider the  $p_T$ distributions of $\mc{O}(\lambda^2)$, we can suppress 
$\blpu{p}_s$ in $\delta^{(2)}(\blpu{p}_1+\blpu{p}_2+\blpu{p}_s)$ since $p_{1,2}^{\perp} \sim \mathcal{O} (\lambda)$. 
Then the scattering cross section becomes
\begin{eqnarray}\label{xsec}
\sigma &=& \sigma_0 H(Q^2) \int dp_1^+dp_1^- dp_2^+dp_2^- \frac{(p_1^-)^3 p_2^+}{Q^2} J_n (p_1^2) J_{\n}(p_2^2) \\
&&\times \sum_{X_s} \delta(Q-p_1^--p_2^--p_s^-) \delta(Q-p_2^+-p_1^+-p_s^+) \frac{1}{N_c} \mr{Tr}\langle  0 | 
\tilde{Y}_{\n} \tilde{Y}_n |X_s \rangle \langle 
X_s | \tilde{Y}_n^{\dagger} \tilde{Y}_{\n} | 0 \rangle, \nonumber \\
&=& \sigma_0 H(Q^2) \int dp_1^+dp_1^- dp_2^+dp_2^- dl^+ dl^-\frac{(p_1^-)^3 p_2^+}{Q^2}
\delta(Q-p_1^+-p_2^+-l^+) \delta(Q-p_2^--p_1^--l^-) \nonumber \\ 
&&\times J_n (p_1^2) J_{\n}(p_2^2) \sum_{X_s}  \frac{1}{N_c} \mr{Tr}\langle  0 | \tilde{Y}_{\n} \tilde{Y}_n |X_s \rangle \langle X_s | 
\delta(l^++\mP_s^+) \delta(l^-+\mP_s^-)\tilde{Y}_n^{\dagger} \tilde{Y}_{\n} | 0 \rangle, \nonumber\\
&=& \sigma_0 H(Q^2,\mu) \int dp_2^-  dp_1^+ p_1^- p_2^+
  J_n (p_1^2, \mu) J_{\overline{n}} (p_2^2,\mu)   \int dl^+ dl^-  S(l^+, l^-,\mu), \nonumber
\end{eqnarray}
where $p_1^+=Q-p_2^+-l^+$, $p_2^-=Q-p_1^--l^-$, $\blpu{p}_1=\blpu{p}_2=0$, and the soft  function $S(l^+,l^-)$ is defined as 
\begin{equation}\label{softf}
S(l^+,l^-) = \frac{1}{N_c} \mr{Tr}\langle  0 | \tilde{Y}_{\n} \tilde{Y}_n\delta(l^++\mP_s^+) \delta(l^-+\mP_s^-)\tilde{Y}_n^{\dagger} \tilde{Y}_{\n} | 0 \rangle. 
\end{equation}

If a jet algorithm is applied, it constrains both the collinear and soft parts. The total dijet scattering cross section can be written as
\begin{eqnarray} \label{facjet} 
\sigma_J &=& \sigma_0 H(Q^2,\mu) \int dp_1^+ dp_2^- Q^2 J_{n,\Theta} (p_1^2)
 J_{\overline{n},\Theta} (p^2_2)  \int dl^+ dl^-  S_{\Theta}(l^+, l^-) \nonumber \\
&=& \sigma_0  H(Q^2,\mu)  \int dp_1^2   J_{n,\Theta} (p_1^2,\mu) \int dp^2_2 J_{\overline{n},\Theta}
 (p^2_2,\mu) \mc{S}_{\Theta} (\mu) \nonumber \\ 
&=&\sigma_0  H(Q^2,\mu)    \mc{J}_{n,\Theta} ( \mu)   \mc{J}_{\overline{n},\Theta} (\mu) \mc{S}_{\Theta} (\mu),  
\end{eqnarray}
where the subscript $\Theta$ means the jet algorithm is implemented. And $\mc{S}_{\Theta}$ is the integrated soft function given by
\begin{equation}
\mc{S}_{\Theta} (\mu) =\int dl^+ dl^- S_{\Theta} (l^+,l^-,\mu).
\end{equation}
If we do not apply jet algorithms, the integrated soft function becomes 1 to all orders in $\alpha_s$ because of the unitarity.
For later use, we call $J_{n,\Theta}(p^2,\mu)$ the unintegrated jet function, and 
$\mc{J}_{n,\Theta}(\mu)=\int dp^2 J_{n,\Theta}(p^2, \mu)$ 
the integrated jet function. When we focus on the dijet total scattering cross sections without specific observables, the factorization theorem in Eq.~(\ref{facjet}) is enough.

As clearly seen in Eq.~(\ref{facjet}), the scattering cross section is factorized into the hard, collinear and soft parts. 
And since the jet cross section is a 
physical observable, it should be independent of the renormalization scale $\mu$. 
Also the double differential scattering cross section
\begin{equation} \label{findif}
\frac{d\sigma_J}{dp_1^2 dp^2_2} = \sigma_0  H(Q^2,\mu)   J_{n,\Theta} (p_1^2,\mu) 
 J_{\overline{n},\Theta} (p^2_2,\mu) \mc{S}_{\Theta} (\mu) 
\end{equation}
should be also independent of $\mu$. That is, the integrated jet function and the unintegrated jet function have the same anomalous dimensions.  

Note that the double differential scattering cross section is not physical since $p_1^2$ and $p_2^2$ are not physical observables, 
but it can be used as an intermediate step to obtain the total jet 
cross section. However, 
$p_1^2$, $p^2_2$  in Eq.~(\ref{findif}) are not the measured invariant 
 jet mass squared with the corresponding jet algorithm. The physical invariant jet mass comes from the 
sum of the momenta of the collinear and soft particles inside a jet. The invariant masses squared for the
$n$ and $\overline{n}$ jets when they contain a soft gluon are given as
\begin{eqnarray}
m_{j_1}^2 &=& (p_1 +l)^2  = p_1^2 +Q l_++\mc{O}(\lambda^3), \nonumber \\
m_{j_2}^2 &=& (p_2 +l)^2  = p_2^2 +Q l_-+\mc{O}(\lambda^3). 
\end{eqnarray}
And the double  differential scattering cross section with respect to these physical invariant jet masses is obtained from Eq.~(\ref{xsec}) as 
\begin{eqnarray} \label{twojet}
\frac{d\sigma_J}{dm_{j_1}^2 dm_{j_2}^2} &=& \sigma_0  H(Q^2,\mu)  \int dp_1^2 dp_2^2 dl^+ dl^- 
  J_{n,\Theta} (p_1^2,\mu)
 J_{\overline{n},\Theta} (p_2^2,\mu)    S_{\Theta}(l^+, l^-)  \\
&\times& \delta \Bigl( m_{j_1}^2 - p_1^2 -Q l^+\Bigr) \delta \Bigl( m_{j_2}^2 -p_2^2 -Ql^-\Bigr)
 \nonumber \\
&=& \sigma_0 H(Q^2,\mu) \int dl^+ dl^- J_{n,\Theta}  (m_{j_1}^2-Ql^+, \mu )
 J_{\overline{n},\Theta}  (m_{j_2}^2 -Q l^-,\mu)    S_{\Theta}(l^+, l^-,\mu).  \nonumber
\end{eqnarray}
This factorization theorem is similar to the result  in Ref.~\cite{Fleming:2007qr}, where the double differential scattering cross section for top jet mass
has been studied. 

And the single differential scattering cross section is given as
\begin{eqnarray} \label{onejet}
\frac{d\sigma_J}{dm_{j_1}^2} &=& \sigma_0  H(Q^2,\mu)  \int dp_1^2 dp_2^2 dl^+ dl^- 
  J_{n,\Theta} (p_1^2,\mu)
 J_{\overline{n},\Theta} (p_2^2,\mu) S_{\Theta}(l^+, l^-) \delta \Bigl( m_{j_1}^2 - p_1^2 -Q l^+\Bigr)\nonumber \\
&=& \sigma_0 H(Q^2,\mu) \int dl^+ J_{n,\Theta}  (m_{j_1}^2-Ql^+, \mu ) S_{\Theta}(l^+,\mu)
 \mc{J}_{\overline{n},\Theta}  (\mu),
\end{eqnarray}
where $S_{\Theta}(l^+,\mu)$ is given by 
\be\label{sl-}
S_{\Theta} (l^+,\mu) = \int dl^- S_{\Theta}(l^+,l^-,\mu).
\ee

If we are interested in the dependence of the invariant jet masses, 
Eq.~(\ref{twojet}) or (\ref{onejet}) should be used. They are given by the convolution of
the jet and the soft functions. But if we are interested in the total
jet cross section, Eqs.~(\ref{facjet}) and (\ref{findif}) can also be used. 
In order to obtain the integrated and unintegrated jet functions, Eqs.~(\ref{facjet}) and (\ref{findif}) are employed from now on. 
Note that the total dijet cross section in Eq.~(\ref{facjet}) is a product of of the integrated jet functions and the soft function.

\section{Jet algorithms}
Among many jet algorithms, here we describe a generic cone-type and the SW jet algorithms. The characteristics of other jet algorithms will be treated
elsewhere. The basic idea of a jet algorithm is to combine final-state particles into a jet if those particles 
are within the radius of the jet cone $R$ (cone-type algorithm) or the cone of half-angle $\delta$ (SW algorithm). 
The main difference lies in the fact that the cone-type algorithm is concerned about the angle of a particle with respect
to a specific axis such as the thrust axis of the jet, while the SW algorithm is concerned about the relative
angle between the particles  to combine them into a jet.
Since the algorithms are similar conceptually in the sense that   particles form a jet inside a jet cone, 
they show similar behavior in the jet  and the  soft functions.  

Let us be more quantitative on the jet algorithm. We consider at the next-to-leading order, and there are two particles at most in a jet. In $e^+ e^-$ 
annihilation, 
a $q\overline{q}$ pair will be produced at leading order forming a back-to-back jet. A gluon can be emitted at order $\alpha_s$, and it can
be combined with a quark or an antiquark to form a jet. A $q\overline{q}$ pair can also form a jet, but its contribution to the jet cross section 
 is suppressed compared to the former cases.

\begin{figure}[t] 
\begin{center}
\includegraphics[width=6cm]{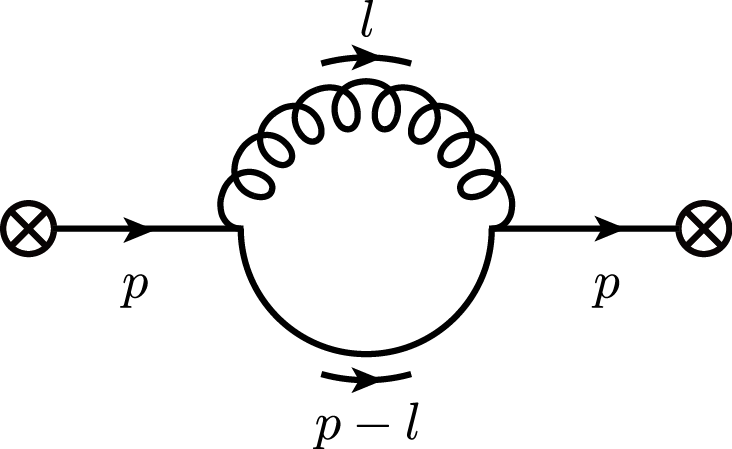}
\end{center}  
\vspace{-0.3cm}
\caption{A schematic diagram for the jet function at next-to-leading order. The jet momentum is $p$, and particles in the loop
carry momentum $l$ (gluon) and $p-l$ (quark).\label{conf}}
\end{figure}

We show schematically the assignment of momenta in Fig.~\ref{conf}. The total momentum of the final state is $p$ with $p_- = Q$ for the jet in the $n$ direction. And $p^2$ is the invariant-mass squared of the collinear
jet. In order to obtain the scattering cross section, we cut the diagram in all possible ways and add their contributions. 
When we cut the loop, there are two final-state particles with 
the momenta $l$ (for a gluon) and $p-l$ (for a quark or an antiquark).  If we cut a single line, it gives
virtual corrections. 
The collinear gluon momentum $l^{\mu}$ in the $n$ direction scales as
\begin{equation} \label{comom}
l^{\mu} = (l_-,  l_{\perp}, l_+) \sim Q(1,\lambda,\lambda^2),
\end{equation}
where  $Q = p_-$ is the largest momentum component of the collinear jet.

A cone-type algorithm collects all the particles within a fixed cone of radius $R$, which determines the jet size. The jet axis is determined, for 
example, by a thrust axis. There are many issues on how
to construct cone jets including infrared safety. But if we confine ourselves to the jets at the next-to-leading order, there are at most two particles 
in a jet and all the cone-type algorithms are reduced to a generic cone-type algorithm.  For collinear particles, if two particles are in 
the cone of radius $R$, that is, if
$\theta_q <R$, and $\theta_g <R$, where $\theta_q$ and $\theta_g$ are the angles of the quark and the gluon with respect to the jet axis, 
they form a jet. In terms of the momenta  shown in Fig.~\ref{conf}, this corresponds to the cone algorithm for the collinear particles:
\begin{equation}
\Theta_{\mathrm{cone}} = \Theta \Bigl( t^2  > \frac{l_+}{l_-}\Bigr) \Theta \Bigl( t^2   >\frac{s/Q -l_+}{Q-l_-}
\Bigr),
\end{equation}
where  $t = \tan R/2$ with the jet cone radius $R$. Each term in $\Theta_{\mathrm{cone}}$ represents $\theta_g <R$ and $\theta_q <R$ respectively.

In the SW jet algorithm, a two-jet event is defined as the one in which the total energy $Q$ except the fraction $\beta$ is deposited in two jet cones 
with half angle $\delta$. This algorithm can be split into a collinear part and a soft part, according to which the factorized result is written in 
Eq.~(\ref{facjet}). For the collinear part, the two particles are energetic and, if we let $\theta$ be the angle between the two particles, the jet algorithm
states that $\theta < 2\delta$,
which is equivalent to $1-\cos \theta < 1-\cos 2\delta$ for $0<\theta <\pi$. For the jet in the $n$-direction, 
combining the on-shll conditions $l^2 =0$, 
$(p-l)^2=0$,   and using the hierarchy of lightcone momenta, this corresponds to
\begin{equation} \label{cojet}
\Theta_{\mathrm{SW}} = \Theta \Bigl(t^2 >\frac{Q^2 l_+}{l_- (Q-l_-)^2 } \Bigr),
\end{equation}
with $t=\delta$. 

We will use $t$ both for the cone radius ($t=\tan R/2$), and the angle ($t=\delta$).  So far, the cone size is independent of the
power counting of momenta in SCET.   For example, if $R= \pi$ for the entire hemisphere, 
$t = \tan R/2\sim \mathcal{O} (1)$. However, if $t$ is too small, the constraint of the jet algorithm does not work. 
The size of the parameters $t$ and $\beta$, related to the jet algorithm, should be determined by the physics
we consider. As can be seen in Eq.~(\ref{cojet}), 
with the fact that the collinear momentum scales as $(l_-, l_{\perp}, l_+) \sim Q(1,\lambda,\lambda^2)$, the power counting $t^2 \sim\lambda^2$
produces a nontrivial jet algorithm.  Therefore here we consider the case with $t\sim \lambda$,
and $t$ can be used interchangeably with  $\tan R/2$ or $\delta$.  
For the jet in the $\overline{n}$ direction, the corresponding condition is given by switching $l_+$ and $l_-$. 

\begin{figure}[b] 
\begin{center}
\includegraphics[width=16cm]{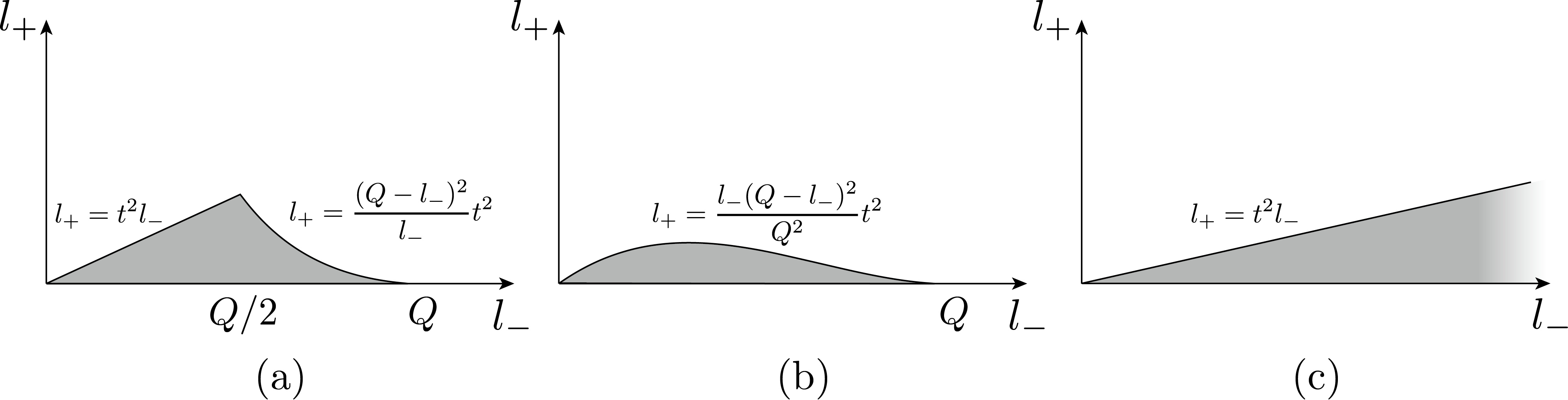}
\end{center}  
\vspace{-0.3cm}
\caption{Constraints on the phase space of the collinear jet function. The shaded regions are allowed by the jet algorithms. (a) Phase space
in the cone-type jet algorithm, (b) Phase space in the SW jet algorithm, (c) Zero-bin region for both jet algorithms. \label{colphase}}
\end{figure}

To avoid double counting, the zero-bin contribution should be considered \cite{Manohar:2006nz},  where 
the collinear momentum $l^{\mu}$ approaches the soft-momentum limit
\begin{equation}
l^{\mu} = (l_+,  l_{\perp}, l_-) \sim Q(\lambda^2,\lambda^2,\lambda^2).
\end{equation}
The corresponding zero-bin phase space is expressed as
\begin{equation}
\Theta^{(0)} = \Theta \Bigl( t^2 > \frac{l_+}{l_-}\Bigr). 
\end{equation}
This is common to the  cone-type and the SW jet algorithms.
The phase spaces given by the jet algorithms are shown in Fig.~\ref{colphase}. Fig.~\ref{colphase} (a) is the phase space available for the naive
collinear contribution in the cone-type jet algorithm, Fig.~\ref{colphase} (b) is the phase space available for the SW jet algorithm, and  
Fig.~\ref{colphase} (c) is the phase space for the zero-bin contribution for both jet algorithms.

We use the same constraints for soft particles in both algorithms. That is, when a soft gluon is inside the cone either in the $n$ or $\overline{n}$
directions, it is included in the jet. An additional constraint on the soft function is about a jet veto. If a soft particle is outside the jet, 
the energy of the soft gluon is less than $\beta$. The soft momentum scales as $Q\lambda^2$, and the nontrivial constraint from the jet veto occurs when
$\beta \sim \lambda^2$. From now on, we fix the power counting on the parameters in the jet algorithm as $t\sim \lambda$ and $\beta \sim 
\lambda^2$.

\begin{figure}[t] 
\begin{center}
\includegraphics[width=5.6cm]{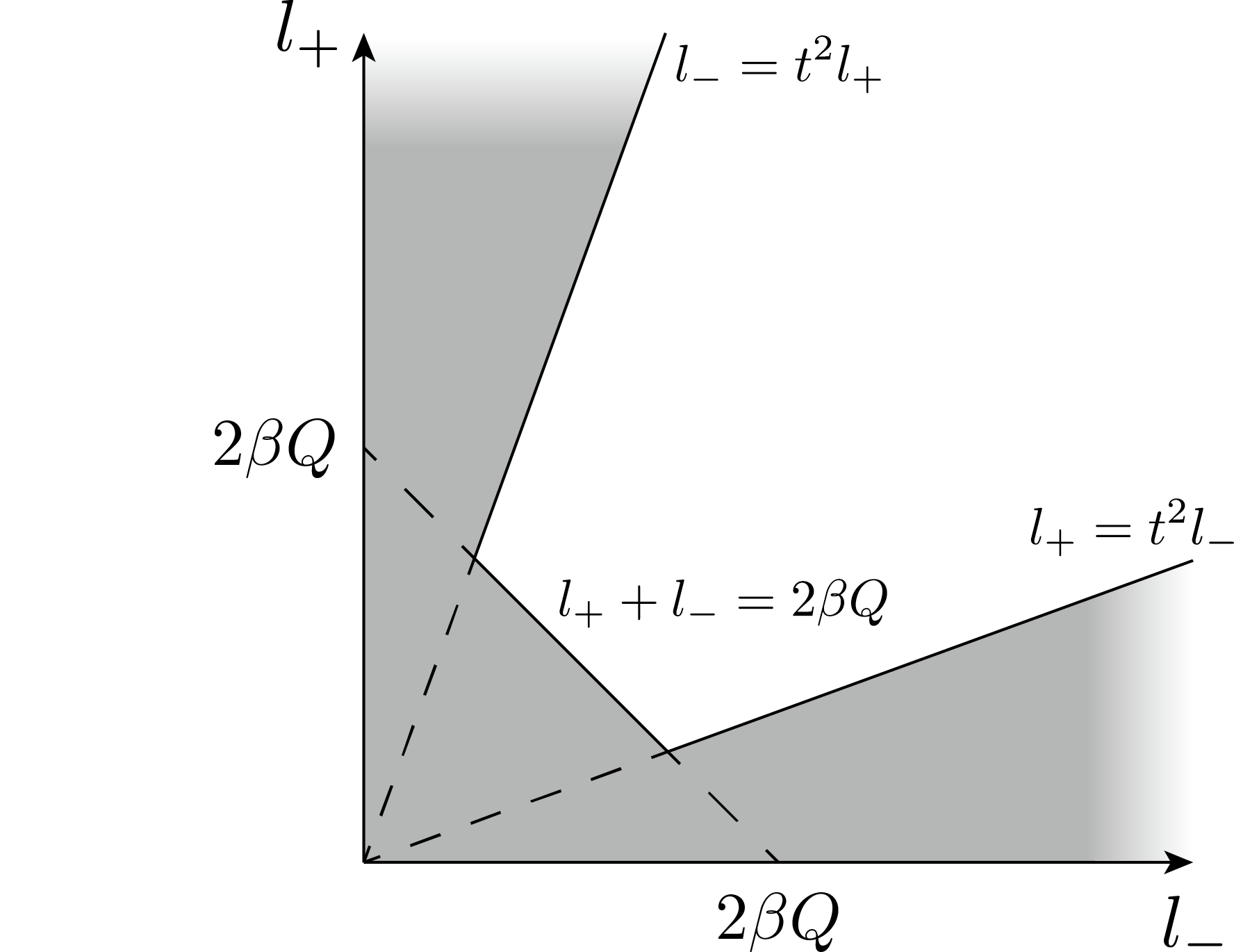}
\end{center}  
\vspace{-0.3cm}
\caption{Phase space available for the soft function. \label{softps}}
\end{figure}

Then the constraint on the soft function from the jet algorithm can be expressed as
\begin{equation} \label{jetsoft}
\Theta_{\mathrm{soft}} =  \left\{ \begin{array}{ll}
\displaystyle \Theta \Bigl( t^2 >\frac{l_+}{l_-}\Bigr), & (n \ \mathrm{jet}), \\
 \displaystyle \Theta \Bigl( t^2 > \frac{l_-}{l_+}\Bigr), & (\overline{n} \ \mathrm{jet}), \\
 \Theta (l_+ + l_- <2\beta Q), & (\mathrm{outside} \ \mathrm{the\ jet}).
\end{array}
\right.
\end{equation}
Fig.~\ref{softps} shows the phase space for the soft function. The shaded area is allowed by the jet algorithm in Eq.~(\ref{jetsoft})

Note that we include the constraint from the jet veto in the jet algorithm of the soft function, not in the collinear function. This is obvious from the 
power counting since $\beta \sim \lambda^2$ and $l_+ +l_- \sim Q$ for the collinear gluon, hence there is no veto needed 
for the collinear particles. Instead the zero-bin 
subtraction eliminates the soft limit in the collinear jet function.  However, in Ref.~\cite{Jouttenus:2009ns}, the author considered only the jet
function with the same power counting on $t$ and $\beta$, but this jet veto was also included in the jet function. 
This seems to be a different definition of the jet function, compared to ours. But the jet function in Ref.~\cite{Jouttenus:2009ns} does not depend on
$\beta$ after the zero-bin subtraction. That is, the inclusion of a jet veto in the jet function does not alter the result.
However,  we stress that our definition of the jet algorithm strictly respects the specified power counting of the parameters.

A question arises if the cone-type and the SW jet algorithms can also be employed at hadron colliders. The main difference between $e^+ e^-$ 
annihilation and $pp$ scattering is that the center of energy is fixed in $e^+ e^-$ 
annihilation, while it is not fixed in  $pp$ scattering. And the center-of-energy frames at the partonic level in $pp$ scattering have a boost invariance
along the beam direction. Therefore the jet algorithm in $pp$ scattering should be defined in a boost-invariant way, for example, in terms of
the rapidity and the azimuthal angle. The cone-type algorithm can be employed in $pp$ scattering, and the method of defining the jet axis should 
be employed.  The SW jet algorithm seems to be specific to $e^+ e^-$ annihilation. 
However, at the next-to-leading order in which there are three final-state particles, the inclusive $k_T$ algorithms ($k_T$, Cambridge/Aachen, 
and  anti-$k_T$) reduce to the same constraint  as the SW jet algorithm, and can be used in hadron collisions.  

\section{Integrated jet function}
The unintegrated jet function $J_{n,\Theta}  (p^2,\mu) $ in the $n$ direction is defined as
\begin{equation}
\sum_{X_n} \langle 0| \chin^{\alpha} |X_n\rangle \Theta_J \langle X_n | \overline{\chi}_n^{\beta}|0\rangle =
\int \frac{d^4 p_{X_n}}{(2\pi)^3} \frac{\FMslash{n}}{2} \overline{n} \cdot p_{X_n} J_{n,\Theta}  (p_{X_n}^2,\mu) \delta^{\alpha\beta},
\end{equation}
where  $\Theta_J$ is the jet algorithm.
The integrated jet function $\mathcal{J}_{n,\Theta}  (\mu)$ is defined as
\begin{equation} \label{inunin}
\mathcal{J}_{n,\Theta}  (\mu)= \int dp^2 J_{n,\Theta} (p^2,\mu),
\end{equation}
where $p^2$ is the invariant mass squared of the collinear jet. 
The unintegrated jet function is the collinear jet function with $p^2$ fixed, and it is expressed in terms of distribution functions. When integrated over
$p^2$, the integrated jet function is obtained.
Eq.~(\ref{inunin})
should hold to all orders in $\alpha_s$. But in current literature, there is some confusion between the integrated jet function, which is obtained by 
integrating the unintegrated jet function, and the inclusive jet function in which there is no jet algorithm involved.  

In a different context in Ref.~\cite{Ellis:2010rwa}, the 
unmeasured jet function (which is the same as the integrated jet function in our case) is not obtained from the measured jet function (which corresponds
to the unintegrated jet function, but different)   by integrating with respect to the thrust. The authors claim that the small parameter in 
power counting is different. That is, in a measured jet sector the small parameter is $\sqrt{\tau_a} \sim \lambda$, while $R\sim \lambda^0$. Indeed different power counting yields different answers, but as far as the total jet cross section is concerned, the result through the 
paths from the integrated jet functions, and from the measured jet functions is the same. The relation between these two approaches and the physics
behind them will be clarified in Section VIII. 

In Sections IV and V, we explicitly compute the integrated and the unintegrated jet functions separately. 
The difference lies in the different order of integration of the delta functions, but there are subtleties which 
arise in the detailed calculations.  What we also focus on is whether the factorized parts, the jet function and the soft function, are infrared finite  themselves. If they contain 
infrared divergences though the sum of the jet and the soft functions do not, the factorized result is not physically meaningful.
In order to show the IR safety explicitly,  we employ the dimensional regularization with $D=4-2\eps$ and the $\overline{\mathrm{MS}}$ scheme, 
and regulate both the UV and the IR divergences. In the $\overline{\mathrm{MS}}$ scheme, we put $4\pi \mu_{\overline{\mathrm{MS}}}^2
= \mu^2 e^{\gamma_{\mathrm{E}}}$.

We can also employ 
the dimensional regularization for the UV divergence and the rapidity regulator only 
in the collinear Wilson line \cite{Chay:2012mh, Chay:2013zya} to regulate the rapidity divergence as
\begin{equation} \label{irreg}
W_n =\sum_{\mathrm{perm}} \exp \Bigl[ -\frac{g}{\overline{n}\cdot \mathcal{P} +\Delta +i0} \overline{n}\cdot A_n\Bigr].
\end{equation}
The final collinear results with the zero-bin subtraction are independent of the regulator $\Delta$ though it appears
in individual diagrams, and the result is the same as that with pure dimensional regularization. We present the result
for the integrated jet function in the SW jet algorithm in Appendix for completeness.
  
\begin{figure}[t] 
\begin{center}
\includegraphics[width=16cm]{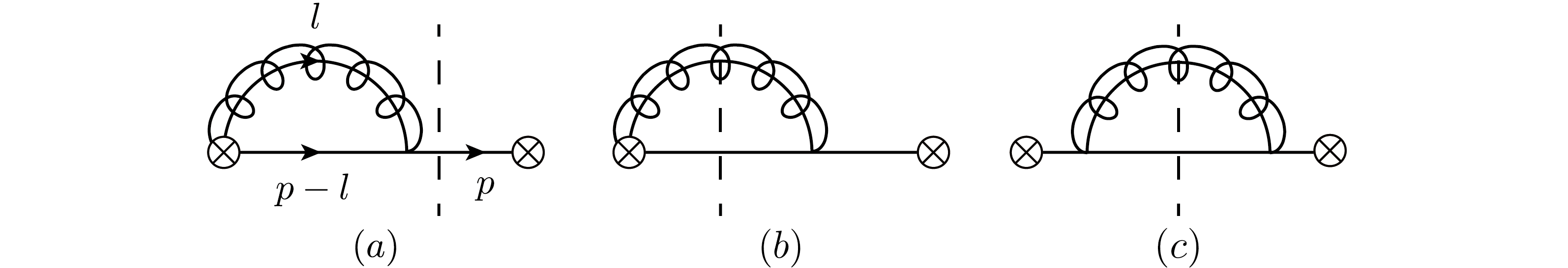}
\end{center}  
\vspace{-0.3cm}
\caption{Feynman diagrams for the jet  functions at one loop.  
(a) virtual correction, (b--c) real gluon emission. The mirror images of (a) and (b) are omitted.\label{intjet}}
\end{figure}

The Feynman diagrams for the jet function, say, in the $n$ direction, are shown in Fig.~\ref{intjet}. The dashed line represents the cut, therefore
 Fig.~\ref{intjet} (a) is the virtual correction, while Fig.~\ref{intjet} (b) and (c) are the contributions from the real gluon emission. 
The mirror images of (a) and (b)  are omitted in Fig.~\ref{intjet}, but they are included in the computation. And we also include the wave function
renormalization and the residue from the self-energy diagram for the quark. Each diagram is accompanied by the 
corresponding zero-bin contribution, which should be subtracted to obtain the collinear jet function.   

\subsection{Cone-type jet algorithm}
The naive collinear virtual correction in Fig.~\ref{intjet} (a) is given by
\begin{equation}
\tilde{M}_a = 2ig^2 C_F   \Bigl(\frac{\mu^2 e^{\gamma_{\mathrm{E}}}}{4\pi}\Bigr)^{\eps} \int \frac{d^D l}{(2\pi)^D} 
\frac{Q-l_-}{l^2 (p-l)^2 l_-} 
=\frac{\alpha_s C_F}{2\pi} 
\Bigl( \frac{1}{\euv} -\frac{1}{\eir}\Bigr) \Bigl( \frac{1}{\eir} +1 +\ln \frac{\mu}{Q}\Bigr) ,
\end{equation}
where $Q=p_-$ is the largest component of the lightcone momentum in the $n$ direction.
In obtaining the final result, we use the fact that
\begin{equation} \label{dimcl}
\mu^{\eps} \int_0^{\infty} du u^{-1-\eps}  =\frac{1}{\euv}-\frac{1}{\eir},
\end{equation}
where $u$ is the integration variable with the dimension of mass. 
In the dimensional regularization in which the UV and the IR poles are not distinguished, the above integral is simply zero because it is a scaleless
integral. However, if we regulate
the IR divergence by inserting a regulator as
\begin{equation} \label{dimreg}
\mu^{\eps} \int_0^{\infty} du (u+\delta)^{-1-\eps}  =\frac{1}{\euv}\Bigl(\frac{\mu}{\delta}\Bigr)^{\eps} = \frac{1}{\euv} - \ln \frac{\mu}{\delta},
\end{equation}
the UV pole can be explicitly extracted. Therefore in order for the integral in Eq.~(\ref{dimcl}) without the regulator  to satisfy the requirement that
it be zero, the pole structure should be of the form given in Eq.~(\ref{dimcl}).  Note that we presume $\euv >0$ to assert that the integrand at infinity
vanishes in Eq.~(\ref{dimreg}). For the same reason, we presume that $\eir <0$. 

The corresponding zero-bin contribution is given as
\begin{equation}
M_a^0 =-\frac{\alpha_s C_F}{2\pi} \Bigl( \frac{1}{\euv} -\frac{1}{\eir}\Bigr)^2.
\end{equation}
The collinear contribution from Fig.~\ref{intjet} (a) is given by
\begin{equation} \label{cola}
M_a = \tilde{M}_a -M_a^0 = \frac{\alpha_s C_F}{2\pi}  \Bigl( \frac{1}{\euv} -\frac{1}{\eir}\Bigr)  \Bigl( \frac{1}{\euv} +1+\ln \frac{\mu}{Q}\Bigr).
\end{equation}
Since the virtual corrections are independent of the jet algorithm at order $\alpha_s$, this result is also true in 
the SW algorithm to be described below.

The naive collinear real gluon emission in Fig.~\ref{intjet} (b)  is affected by the jet algorithm and is given  as
\begin{eqnarray} \label{naivemb}
\tilde{M}_b &=& \frac{\alpha_s C_F}{2\pi} \frac{(e^{\gamma_{\mathrm{E}}} \mu^2 )^{\eps}}{\Gamma (1-\eps)} \int dl_+ dl_- 
\frac{Q-l_-}{Q} (l_-  l_+)^{-1-\eps} \Theta_{\mathrm{cone}}  \\
&=&\frac{\alpha_s C_F}{2\pi} \frac{(e^{\gamma_{\mathrm{E}}} \mu^2)^{\eps}}{\Gamma(1-\eps)} \Bigl[ \int_0^{Q/2} dl_- 
\frac{Q-l_-}{Q} l_-^{-1-\eps}\int_0^{t^2l_-} dl_+ l_+^{-1-\eps} \nonumber \\
&&+ \int_{Q/2}^Q dl_- \frac{Q-l_-}{Q} l^{-1-\eps}\int_0^{(Q-l_-)^2 t^2/l_-} dl_+ l_+^{-1-\eps}\Bigr] \nonumber \\
&=& \frac{\alpha_s C_F}{2\pi}  \Bigl[  \frac{1}{2\eir^2} +\frac{1}{\eir} \Bigl(1+ \frac{1}{2}\ln \frac{\mu^2}{Q^2 t^2} \Bigr) + 
\ln \frac{\mu^2}{Q^2 t^2} +\frac{1}{4} \ln^2 \frac{\mu^2}{Q^2 t^2} +2+2\ln 2 -\frac{5}{24}\pi^2\Bigr]. \nonumber
\end{eqnarray}
As can be seen clearly in the integral, there are only IR poles in this naive collinear contribution.

 And the corresponding zero-bin contribution is given as
\begin{eqnarray}
M_b^0 &=& \frac{\alpha_s C_F}{2\pi} \frac{(e^{\gamma_{\mathrm{E}}} \mu^2 )^{\eps}}{\Gamma (1-\eps)} \int dl_+ dl_- 
 (l_- l_+)^{-1-\eps} \Theta^{(0)} \nonumber \\
&=& \frac{\alpha_s C_F}{2\pi} \frac{(e^{\gamma_{\mathrm{E}}} \mu^2 )^{\eps}}{\Gamma (1-\eps)} \int_0^{\infty} dl_- 
 l_-^{-1-\eps}\int_0^{t^2l_- } dl_+ l_+^{-1-\eps}.
\end{eqnarray}
Care must be taken in computing this integral in dimensional regularization. If we naively perform the $l_+$ integral, it yields 
$-(t^2 l_-)^{-\eps}/\eps$, with the presumption that $\eps <0$ to guarantee that the integral vanishes at $l_+=0$. That is, the divergence is of the 
IR origin. This prescription is fine for the $l_-$ integral near $l_- \sim 0$. However, when we perform the $l_-$ integral, 
the integral should also be regulated as $l_-$ approaches infinity. 
It means $\eps >0$ in this integral and the presumption that $\eps <0$ in the $l_+$ integral is violated as $L_-$ approaches infinity. Therefore
we cannot obtain a meaningful answer if we naively compute the integral as it is. 

The trick to avoid this problem is to write the integral as 
\begin{eqnarray} \label{trick}
&&\int_0^{\infty} dl_-  l_-^{-1-\eps}\int_0^{t^2l_- } dl_+ l_+^{-1-\eps} = \Bigl( \int_0^{\eta}dl_- l_-^{-1-\eps} +\int_{\eta}^{\infty} dl_-
l_-^{-1-\eps}\Bigr) \int_0^{t^2l_- } dl_+ l_+^{-1-\eps} \nonumber \\
&&= \int_0^{\eta} dl_- l_-^{-1-\eps} \int_0^{t^2 l_-} dl_+ l_+^{-1-\eps} + \int_{\eta}^{\infty} dl_- l_-^{-1-\eps} \Bigl( \int_0^{\infty} dl_+ 
l_+^{-1-\eps} -\int_{t^2 l_-}^{\infty} dl_+ l_+^{-1-\eps}\Bigr),
\end{eqnarray}
where $\eta$ is some positive energy.  In the first line, the first integral on the right-hand side yields the IR poles ($\eps<0$). The second integral,
as it is, is inconsistent because the $l_+$ integral forces $\eps<0$, while the integral on $l_-$ requires $\eps>0$ as $l_-\rightarrow \infty$. 
Therefore the second $l_+$ integral
is decomposed to yield the final result. The last integral is purely of the UV origin, while the second integral has the UV origin from the $l_-$ integral, 
and the $l_+$ integral can be treated by Eq.~(\ref{dimcl}).   With this careful
manipulation of separating the UV and IR regions, the final
result is independent of the arbitrary scale $\eta$, and the zero-bin contribution is given by
\begin{equation} \label{zerob}
M_b^0 = \frac{\alpha_s C_F}{2\pi}  \Bigl[  \frac{1}{2}\Bigl( \frac{1}{\euv} -\frac{1}{\eir}\Bigr)^2 
+\frac{1}{2}\Bigl( \frac{1}{\euv} -\frac{1}{\eir}\Bigr) \ln t^2\Bigr].
\end{equation}
If this trick were not used and if the integral were computed carelessly, the result would be totally different and we get a meaningless jet function. 
The collinear  contribution from Fig.~\ref{intjet} (b) is finally given by
\begin{eqnarray}
M_b &=& \tilde{M}_b - M_b^0 
= \frac{\alpha_s C_F}{2\pi}  \Bigl[ -\frac{1}{2\euv^2} +\frac{1}{\euv\eir}  +\frac{1}{\eir}\Bigl( 1+\ln \frac{\mu}{Q}\Bigr) 
-\frac{1}{\euv}\ln t  \nonumber \\
&& +\ln \frac{\mu^2}{Q^2 t^2} +\frac{1}{4} \ln^2 \frac{\mu^2}{Q^2 t^2} +2+2\ln 2 -\frac{5}{24}\pi^2\Bigr].  
\end{eqnarray}

The naive collinear contribution shown in Fig.~\ref{intjet} (c) is given as
\begin{eqnarray}
\tilde{M}_c &=& \frac{\alpha_s C_F}{2\pi} (1-\eps)\frac{(e^{\gamma_{\mathrm{E}}} \mu^2)^{\eps}}{\Gamma(1-\eps)} 
\int  dl_+ dl_-     \frac{l_-^{1-\eps}}{Q^2} l_+^{-1-\eps}  
 \Theta_{\mathrm{cone}}   \\
&=&  \frac{\alpha_s C_F}{2\pi} (1-\eps)\frac{(e^{\gamma_{\mathrm{E}}} \mu^2)^{\eps}}{\Gamma(1-\eps)} 
\Bigl[ \int_0^{Q/2} dl_- \frac{l_-^{1-\eps}}{Q^2}\int_0^{t^2l_-} dl_+ l_+^{-1-\eps} 
 + \int_{Q/2}^Q dl_-  \frac{ l_-^{1-\eps}}{Q^2}\int_0^{(Q-l_-)^2 t^2/l_-} dl_+ l_+^{-1-\eps}\Bigr]  \nonumber \\ 
&=& \frac{\alpha_s C_F}{4\pi}\Bigl( -\frac{1}{\eir} -\ln \frac{\mu^2}{Q^2 t^2} -1-2\ln 2 \Bigr). \nonumber 
\end{eqnarray}
Note also that the pole is of the IR origin. The zero-bin contribution is  suppressed at this order. 
 
Finally, the wave function renormalization and the residue from the self-energy correction of the fermion at order $\alpha_s$ are given as
\begin{equation}
Z_{\xi}^{(1)} +R_{\xi}^{(1)} = -\frac{\alpha_s C_F}{4\pi} \Bigl( \frac{1}{\euv}-\frac{1}{\eir}\Bigr). 
\end{equation}
Combining all the contributions by including proper mirror images, the total contribution to the integrated jet function  
is given by
\begin{eqnarray} \label{conejet}
M_{\mathrm{coll}} &=& 2(M_a +M_b) +M_c +Z_{\xi}^{(1)} +R_{\xi}^{(1)}  \\
&=& \frac{\alpha_s C_F}{2\pi} \Bigl[ \frac{1}{\euv^2}  +\frac{1}{\euv} \Bigl( \frac{3}{2} +\ln \frac{\mu^2}{Q^2 t^2} 
\Bigr)
+\frac{3}{2} \ln \frac{\mu^2}{Q^2 t^2} +\frac{1}{2} \ln^2 \frac{\mu^2}{Q^2 t^2} +\frac{7}{2} +3\ln 2 -\frac{5\pi^2}{12}\Bigr]. \nonumber 
\end{eqnarray}
Note that the integrated jet function in the cone-type algorithm is indeed IR finite. By removing the UV poles, the integrated jet function in the 
cone-type jet algorithm at order $\alpha_s$ is given by
\begin{equation}
\mathcal{J}_{\mathrm{cone}}^{(1)} (Q,t,\mu) = \frac{\alpha_s C_F}{2\pi} \Bigl( 
\frac{3}{2} \ln \frac{\mu^2}{Q^2 t^2} +\frac{1}{2} \ln^2 \frac{\mu^2}{Q^2 t^2} +\frac{7}{2} +3\ln 2 -\frac{5\pi^2}{12}\Bigr).
\end{equation}
This coincides with the unmeasured quark jet function in Ref.~\cite{Ellis:2010rwa}.

\subsection{SW algorithm}

Since the SW algorithm and the cone-type algorithm are essentially the same except the fact that the cone is determined either by the relative angle of
the two partons, or by the angle with respect to a fixed jet axis such as the thrust axis, the calculation for the jet function in the SW jet algorithm is 
very similar to the jet function in the cone-type jet algorithm except some constants.   

Since the virtual corrections in Figs.~\ref{intjet} (a) and the wave function renormalization are independent of the jet algorithms, 
they are the same as in the case of the cone-type jet  algorithm.  
The naive collinear real gluon emission in Fig.~\ref{intjet} (b)  is affected by the jet algorithm and is given  as
\begin{eqnarray}
\tilde{M}_b &=& \frac{\alpha_s C_F}{2\pi} \frac{(e^{\gamma_{\mathrm{E}}} \mu^2 )^{\eps}}{\Gamma (1-\eps)} \int dl_+ dl_- 
\frac{Q-l_-}{Q} (l_-  l_+)^{-1-\eps} \Theta_{\mathrm{SW}} \nonumber \\
&=& \frac{\alpha_s C_F}{2\pi} \frac{(e^{\gamma_{\mathrm{E}}} \mu^2 )^{\eps}}{\Gamma (1-\eps)} \int_0^Q dl_- 
\frac{Q-l_- }{Q} l_-^{-1-\eps} \int_0^{l_- t^2 (Q-l_-)^2/Q^2} dl_+ l_+^{-1-\eps}\nonumber \\
&=& \frac{\alpha_s C_F}{2\pi}  \Bigl[  \frac{1}{2\eir^2} +\frac{1}{\eir} \Bigl(1+ \frac{1}{2}\ln \frac{\mu^2}{Q^2 t^2} \Bigr) + 
\ln \frac{\mu^2}{Q^2 t^2} +\frac{1}{4} \ln^2 \frac{\mu^2}{Q^2 t^2} +4 -\frac{3}{8}\pi^2\Bigr].
\end{eqnarray}
As can be seen clearly in the integral, there are only IR poles. Since the zero-bin constraint is the same as that
of the cone-type jet algorithm, the zero-bin contribution is also given by Eq.~(\ref{zerob}).
 Finally, the collinear contribution from the real gluon emission is given by
\begin{eqnarray}
M_b &=& \tilde{M}_b - M_b^0 
= \frac{\alpha_s C_F}{2\pi}  \Bigl[ -\frac{1}{2\euv^2} +\frac{1}{\euv\eir}  +\frac{1}{\eir}\Bigl( 1+\ln \frac{\mu}{Q}\Bigr) 
-\frac{1}{\euv}\ln t   \nonumber \\
&& +\ln \frac{\mu^2}{Q^2 t^2} +\frac{1}{4} \ln^2 \frac{\mu^2}{Q^2 t^2} +4-\frac{3}{8}\pi^2\Bigr].
\end{eqnarray}
 
The naive collinear contribution shown in Fig.~\ref{intjet} (c) is given as
\begin{eqnarray}
\tilde{M}_c &=& \frac{\alpha_s C_F}{2\pi} (1-\eps)\frac{(e^{\gamma_{\mathrm{E}}} \mu^2)^{\eps}}{\Gamma(1-\eps)} 
\int  dl_+ dl_-     \frac{l_-^{1-\eps}}{Q^2} l_+^{-1-\eps}  
\Theta_{\mathrm{SW}}   \\
&=&  \frac{\alpha_s C_F}{2\pi} (1-\eps)\frac{(e^{\gamma_{\mathrm{E}}} \mu^2)^{\eps}}{\Gamma(1-\eps)}  
\int_0^Q dl_- \frac{l_-^{-1-\eps}}{Q^2} \int_0^{l_- t^2 (Q-l_-)^2/Q^2} dl_+ l_+^{-1-\eps} \nonumber \\
&=& \frac{\alpha_s C_F}{4\pi}\Bigl( -\frac{1}{\eir} -\ln \frac{\mu^2}{Q^2 t^2} -3 \Bigr). \nonumber 
\end{eqnarray}
Note also that the pole is of the IR origin. The zero-bin contribution is also suppressed at this order. 
 
Adding all the contributions including mirror images of (a) and (b), the total contribution to the integrated jet function for the SW jet 
algorithm is given as
\begin{eqnarray} \label{swjet}
M_{\mathrm{coll}} &=& 2 (M_a +M_b) + M_c + Z_{\xi}^{(1)} +R_{\xi}^{(1)}\\
&=& \frac{\alpha_s C_F}{2\pi} \Bigl[ \frac{1}{\euv^2} +\frac{1}{\euv} \Bigl( \frac{3}{2} + \ln \frac{\mu^2}{Q^2 t^2} \Bigr)
+\frac{3}{2} \ln \frac{\mu^2}{Q^2 t^2} +\frac{1}{2} \ln^2 \frac{\mu^2}{Q^2 t^2} +\frac{13}{2} -\frac{3}{4} \pi^2\Bigr]. \nonumber
\end{eqnarray}
Therefore the integrated jet function in the SW jet algorithm at order $\alpha_s$ is given by
\begin{equation}
\mathcal{J}_{\mathrm{SW}}^{(1)} (Q,t,\mu) = \frac{\alpha_s C_F}{2\pi} \Bigl(  \frac{3}{2} \ln \frac{\mu^2}{Q^2 t^2} 
+\frac{1}{2} \ln^2 \frac{\mu^2}{Q^2 t^2} +\frac{13}{2} -\frac{3}{4} \pi^2\Bigr).
\end{equation}
This is also consistent with the $k_T$-type result in Ref.~\cite{Ellis:2010rwa}, since the SW algorithm and the $k_T$-type algorithm are the same at
this order.
Note that, compared to the contribution to the integrated jet function in the cone-type jet algorithm, all the divergence structure 
and the logarithmic terms are the same. The
only difference comes from the constant terms. This is obvious when we look at the phase space for both jet algorithms  Fig.~\ref{colphase}. 
As $l_-$ goes either to zero
or $Q$, the phase space coincides, which yields the same divergences and logarithms. The difference of the remaining phase space yields different
constant terms. That is why we have stated that the cone-type and the SW jet algorithms are similar. 

\section{Unintegrated jet function}
 
Here we derive the unintegrated jet function for  the cone-type and the SW jet algorithms.
It should yield the integrated jet function when integrated over $p^2$.  
In computing the integrated jet function, we integrated the delta function responsible for the on-shellness of the final-state particles with respect to 
$p^2$ first, and the remaining $l_+$ and $l_-$ integrals were performed. Here we do not integrate over $p^2$. Instead, we perform the $l_+$ and $l_-$
integrals first and express the result in terms of the distribution functions with respect to $p^2$. The integrated jet function should be independent 
of the order of integration, and our calculation explicitly shows that fact. However, there are some caveats on how to deal with 
the distribution function, which will be explained here.
  
\subsection{Unintegrated jet function in the cone-type algorithm}
The Feynman diagrams for the unintegrated jet function are also given by Fig.~\ref{intjet}.  
The virtual correction in Fig.~\ref{intjet} (a) and its zero-bin contribution are the same as the result in the integrated jet function except the
delta function and 
the collinear result is given as
\begin{equation}
M_a = \tilde{M}_a -M_a^0 = \frac{\alpha_s C_F}{2\pi}\delta (p^2)  \Bigl( \frac{1}{\euv} -\frac{1}{\eir}\Bigr)  \Bigl( \frac{1}{\euv} +\ln 
\frac{\mu}{Q}+1\Bigr).
\end{equation}

The naive collinear contribution in Fig.~\ref{intjet} (b) is given as
\begin{eqnarray} \label{unmb}
\tilde{M}_b &=& \frac{\alpha_s C_F}{2\pi} \frac{(e^{\gamma_{\mathrm{E}}} \mu^2)^{\eps}}{\Gamma(1-\eps)}\frac{1}{p^2}\int   dl_+ dl_- 
 \delta ((l-p)^2) (l_+ l_-)^{-\eps} \frac{Q-l_-}{l_- }   \Theta_{\mathrm{cone}}  \nonumber \\
&=&\frac{\alpha_s C_F}{2\pi} \frac{(e^{\gamma_{\mathrm{E}}} \mu^2)^{\eps}}{\Gamma(1-\eps)}\frac{1}{(p^2)^{1+\eps}}
\int_{p^2/w^2}^{Q^2 t^2/w^2} dy y^{-1-\eps} (1-y)^{1-\eps},
\end{eqnarray} 
where $y=l_-/Q$ and $w^2 = Q^2 t^2 +p^2$. Note that $p^2 <Q^2 t^2$ in order for this integral to exist.
 Since $p^2$ also appears in the integration limits, the dependence on $p^2$ does not come only from $(p^2)^{-1-\eps}$ in front of the integral. 
However we observe that Eq.~(\ref{unmb}) is singular at $p^2 =0$, and can be regarded as a distribution function. It can be written as
\begin{equation}
\tilde{M}_b = A\delta (p^2) +[B(p^2)]_{M^2},
\end{equation}
where all the singularities are concentrated at $p^2 =0$ and the remaining part gives a distribution function away from $p^2 =0$, 
which we call the $M^2$-distribution. It is defined as
\begin{equation}
\int_0^{\Lambda^2} dp^2[g(p^2)]_{M^2} f(p^2)  = \int_0^{\Lambda^2}dp^2 g(p^2) f(p^2) -\int_0^{M^2} dp^2 g(p^2) f(0),
\end{equation}
for a well-behaved test function $f(p^2)$. Here  $g(p^2)$ is a function which diverges at $p^2=0$ and $\Lambda$ is an appropriate finite upper limit
for the relevant physical processes in consideration.  It has the following properties:
\begin{eqnarray}
\int_0^{M^2} dp^2[g(p^2)]_{M^2} f(p^2)  &=& \int_0^{M^2}dp^2 g(p^2) \Bigl(f(p^2) -f(0)\Bigr), \nonumber \\
\int_0^{M^2} dp^2[g(p^2)]_{M^2}  &=&0. 
\end{eqnarray}
For the jet algorithms here, we can put $M=Qt$ in the final result since $p^2 \leq Q^2 t^2$. 

 The $M^2$-distribution part can be computed with $\epsilon=0$, and is given as
\begin{equation}
\frac{1}{p^2} \int_{p^2/w^2}^{Q^2 t^2/w^2}  dy \frac{1-y}{y} = \frac{1}{p^2}\Bigl[ -\frac{Q^2 t^2 -p^2}{Q^2 t^2 +p^2} +\ln \frac{Q^2 t^2}{p^2}
\Bigr].
\end{equation}
The part proportional to $\delta (p^2)$ is obtained by integrating
$\tilde{M}_b$ with respect to $p^2$ as†
\begin{equation}
\int_0^{M^2} \frac{dp^2}{(p^2)^{1+\eps}} \int_{p^2/w^2}^{Q^2 t^2/w^2}  dy y^{-1-\eps}(1-y)^{-\eps}.
\end{equation}
Using integration by parts, the calculation is straightforward, and the final result is given as 
\begin{eqnarray}
\tilde{M}_b &=& \frac{\alpha_s C_F}{2\pi} \Bigl\{ \delta (p^2) \Bigl[ \frac{1}{2\eir^2} +\frac{1}{\eir} 
\Bigl(1+\frac{1}{2}\ln \frac{\mu^2}{Q^2 t^2} \Bigr) +\frac{1}{4} \ln^2 \frac{\mu^2}{Q^2 t^2}+2-\frac{5}{24}\pi^2 \nonumber \\
&&+\ln \frac{\mu^2}{M^2} -\frac{1}{2}\ln^2 \frac{Q^2 t^2}{M^2} -2\ln \frac{Q^2 t^2}{Q^2 t^2 +M^2}\Bigr] 
+\Bigl[\frac{1}{p^2}\Bigl( -\frac{Q^2 t^2 -p^2}{Q^2 t^2 +p^2} +\ln \frac{Q^2 t^2}{p^2}\Bigr)\Bigr]_{M^2}\ \Bigr\}.
\end{eqnarray}
 
The zero-bin contribution  is given by
\begin{equation} \label{zerobun}
M_b^0 =  \frac{\alpha_s C_F}{2\pi} \frac{(e^{\gamma_{\mathrm{E}}} \mu^2)^{\eps}}{\Gamma(1-\eps)}\frac{1}{(p^2)^{1+\eps}}
\int_{p^2/Q t^2}^{\infty} \frac{dl_-}{Q} \Bigl( \frac{l_-}{Q}\Bigr)^{-1-\eps}. 
\end{equation}
We have to be careful in expressing this as a distribution function. Note first that the upper limit of the integration is set to infinity.  
In the naive collinear contribution, the collinear momentum $l^{\mu}$ scales as
$(l_-, l_{\perp}, l_+)\sim Q(1,\lambda,\lambda^2)$. In the zero-bin limit, $l^{\mu}$ scales as $(l_-, l_{\perp}, l_+)\sim Q(\lambda^2, \lambda^2,
\lambda^2)$. The upper limit is of order 1 since it lies in the range $1/2 < Q^2 t^2/(Q^2 t^2 +p^2) <1$ for $0<p^2 <Q^2 t^2$, 
while $l_-/Q \sim \mathcal{O}
(\lambda^2)$. Therefore in the effective theory, the upper limit is set to infinity. It also means that $p^2/Q^2$ is at most 
$\mathcal{O} (\lambda^2)$ or smaller. 
If we assume that $p^2$ is of order $Q^2 \lambda^2$ as in the naive collinear contribution, the lower limit of the integral in Eq.~(\ref{unmb}), 
$p^2 /(Q^2 t^2+p^2)$, becomes of order 1 since $t^2 \sim \mathcal{O} (\lambda^2)$, while $l_-/Q \sim \mathcal{O} (\lambda^2)$. In this case,
the lower limit should be set to infinity also, which is not sensible. Therefore, in computing the zero-bin contribution, we should  
consider $p^2$ much smaller than $Q^2 \lambda^2$, say, $Q^2 \lambda^4$. It means that the zero-bin contribution is concentrated near the origin,
and there is no remaining distribution function away from $p^2 =0$ which describes the region $p^2 \sim Q^2 \lambda^2$. 
In summary, the zero-bin contribution is proportional only to $\delta (p^2)$.

With the above argument,  the coefficient of $\delta (p^2)$ in the zero-bin contribution  can be obtained by integrating Eq.~(\ref{zerobun}) with respect
to $p^2$ from 0 to infinity. Here also, the trick similar to Eq.~(\ref{trick}) should be employed to avoid using the wrong sign of $\eps$. By
shuffling the integration region appropriately and introducing an intermediate scale $\eta$, we can obtain the unambiguous UV and IR divergent
terms as
\begin{equation} \label{zerobuna}
M_b^0 = \frac{\alpha_s C_F}{2\pi}  \delta (p^2) \Bigl[  \frac{1}{2}\Bigl( \frac{1}{\euv} -\frac{1}{\eir}\Bigr)^2 
+\frac{1}{2}\Bigl( \frac{1}{\euv} -\frac{1}{\eir}\Bigr) \ln t^2\Bigr].
\end{equation}
Note that the coefficient of $\delta (p^2)$ is the same as the zero-bin contribution, $M_b^0$, in calculating the integrated jet function in 
Eq.~(\ref{zerob}). The net collinear contribution from Fig.~\ref{intjet} (b) is given as
\begin{eqnarray}
M_b &=& \tilde{M}_b -M_b^0\nonumber \\
& &= \frac{\alpha_s C_F}{2\pi}  \Bigl\{ \delta (p^2) \Bigl[ -\frac{1}{2\euv^2} +\frac{1}{\euv \eir} +
\frac{1}{\eir} \Bigl( 1+\ln \frac{\mu}{Q}\Bigr) 
-\frac{1}{\euv}\ln t  \nonumber \\
&& +\ln \frac{\mu^2}{Q^2 t^2} +\frac{1}{4} \ln^2 \frac{\mu^2}{Q^2 t^2} +2+2\ln 2 -\frac{5}{24}\pi^2  
 + \ln \frac{Q^2 t^2}{M^2} -\frac{1}{2} \ln^2 \frac{Q^2 t^2}{M^2} -2 \ln \frac{2Q^2 t^2}{Q^2 t^2 +M^2} \Bigr]  \nonumber \\
&&+\Bigl[ \frac{1}{p^2} \Bigl( -\frac{Q^2 t^2 -p^2}{Q^2 t^2 +p^2} +\ln \frac{Q^2 t^2}{p^2}\Bigr)\Bigr]_{M^2}\Bigr\}.
\end{eqnarray}

The naive collinear contribution from Fig.~\ref{intjet} (c) is given as
\begin{eqnarray}
\tilde{M}_c &=& \frac{\alpha_s C_F}{2\pi} (1-\eps)\frac{(e^{\gamma_{\mathrm{E}}} \mu^2)^{\eps}}{\Gamma(1-\eps)}  \frac{1}{(p^2)^{1+\eps}}
\int_{p^2/w^2}^{Q^2 t^2/w^2} dy(1-y)^{-\eps} y^{1-\eps} \nonumber \\
&=& \frac{\alpha_s C_F}{4\pi} \Bigl\{ \delta (p^2)\Bigl(-\frac{1}{\eir}   - \ln \frac{\mu^2}{Q^2 t^2} -1-2\ln 2   \nonumber \\
&-&\ln  \frac{Q^2 t^2}{M^2} + 2\ln \frac{2Q^2 t^2}{Q^2 t^2 +M^2}  \Bigr) +\Bigl[ \frac{1}{p^2} \frac{Q^2 t^2 -p^2}{Q^2 t^2 +p^2}
\Bigr]_{M^2}\Bigr\}.
\end{eqnarray}
The zero-bin contribution is also suppressed, and is set to zero.

The amplitude for the unintegrated jet function is given by
\begin{eqnarray}
M_n^{\mathrm{unint}} (p^2,\mu) &=& 2(M_a +M_b) +M_c +Z_{\xi}^{(1)} +R_{\xi}^{(1)}  \nonumber \\
&=& \frac{\alpha_s C_F}{2\pi} \Bigl\{ \delta(p^2) \Bigl[  \frac{1}{\euv^2} +\frac{1}{\euv} \Bigl( \frac{3}{2}  
+\ln \frac{\mu^2}{Q^2 t^2}\Bigr) +\frac{3}{2} \ln \frac{\mu^2}{Q^2 t^2} +\frac{1}{2} \ln^2 \frac{\mu^2}{Q^2 t^2} 
\nonumber \\
&&+ \frac{7}{2} +3\ln 2 -\frac{5\pi^2}{12}    + \frac{3}{2} \ln \frac{Q^2 t^2}{M^2} 
-\ln^2 \frac{Q^2 t^2}{M^2} -3 \ln \frac{2Q^2 t^2}{Q^2 t^2 +M^2} \Bigr] \nonumber \\
&&
+\Bigl[\frac{2}{p^2}\ln \frac{Q^2 t^2}{p^2} + \frac{1}{p^2} \Bigl(\frac{3}{2} -\frac{3Q^2 t^2}{p^2+Q^2 t^2}\Bigr)\Bigr]_{M^2}\Bigr\}.
\end{eqnarray}
If we put $M=Qt$, the last line vanishes. Dropping the UV divergence, the final unintegrated jet function at order $\alpha_s$
in the cone-type jet algorithm is given as
\begin{eqnarray}
J_{\mathrm{cone}}^{(1)}  (p^2, Q,t,\mu) 
&=& \frac{\alpha_s C_F}{2\pi} \Bigl\{ \delta(p^2) \Bigl(  \frac{3}{2} \ln \frac{\mu^2}{Q^2 t^2} +\frac{1}{2} \ln^2 \frac{\mu^2}{Q^2 t^2} 
+ \frac{7}{2} +3\ln 2 -\frac{5\pi^2}{12}\Bigr)  \nonumber \\
&& +\Bigl[\frac{2}{p^2}\ln \frac{Q^2 t^2}{p^2} + \frac{1}{p^2} \Bigl(\frac{3}{2} -
\frac{3Q^2 t^2}{p^2+Q^2 t^2}\Bigr)\Bigr]_{M^2=Q^2 t^2}\Bigr\}.
\end{eqnarray}

Note that the coefficient of $\delta (p^2)$ is exactly the same as the integrated jet function, and the integration of the $M^2$ distribution yields zero. 
Therefore we have successfully obtained the result that the integrated jet function is indeed obtained by integrating the unintegrated jet function. 

\subsection{Unintegrated jet function in the Sterman-Weinberg jet algorithm}
We can proceed to compute the unintegrated jet function in the SW jet algorithm with the same idea and method as in the cone-type jet algorithm.
The virtual contributions from Fig.~\ref{intjet} (a) and the self-energy correction are the same. 
The collinear contribution from Fig.~\ref{intjet} (b) is given as
\begin{eqnarray}
M_b &=& \tilde{M}_b - M_b^0 \nonumber \\
 &=& \frac{\alpha_s C_F}{2\pi} \Bigl\{ \delta (p^2) \Bigl[ -\frac{1}{2\euv^2} +\frac{1}{\euv \eir} -\frac{1}{\euv} \ln t +\frac{1}{\eir} \Bigl(1+
\ln \frac{\mu}{Q}\Bigr) \nonumber \\ 
&&+ 4 -\frac{3}{8}\pi^2 +\ln \frac{\mu^2}{Q^2 t^2} +\frac{1}{4} \ln^2 \frac{\mu^2}{Q^2 t^2} + S_1 (M^2) \Bigr] \nonumber \\
&+& \Bigl[ \frac{1}{p^2} \Bigl(-\sqrt{1-4p^2/Q^2 t^2} +\ln \frac{1+\sqrt{1-4p^2/Q^2 t^2}}{1-\sqrt{1-4p^2/Q^2 t^2}}\Bigr) \Bigr]_{M^2} \Bigr\}.
\end{eqnarray}
Here $S_1 (M^2)$ is given by
\begin{eqnarray}
S_1 (M^2) &=& \mathrm{Li}_2 (w_+) -\mathrm{Li}_2 (w_-) -2\sqrt{1- \frac{4M^2}{Q^2 t^2}}+\ln \frac{w_+}{w_-} \nonumber \\
&-&\ln \frac{M^2}{Q^2 t^2} \ln \frac{M^2/Q^2 t^2}{w_+ - M^2/Q^2 t^2}  
+\frac{1}{2} \ln^2 w_- -\frac{1}{2} \ln^2 w_+,
\end{eqnarray}
where $\mathrm{Li}_2 (x)$ is the dilogarithmic function and
\begin{equation}
w_{\pm} = \frac{1}{2}\Bigl(1\pm \sqrt{1-\frac{4M^2}{Q^2 t^2}}\Bigr).
\end{equation}
Later we will choose $M= Qt/2$, which corresponds to $M^2/Q^2 t^2 = 1/4$, $w_{\pm}=1/2$. And $S_1(Q^2 t^2/4) =0$.

Fig.~\ref{intjet} (c) is given as
\begin{equation}
M_c = \frac{\alpha_s C_F}{2\pi} \Bigl\{ \delta (p^2) \Bigl[-\frac{1}{2\eir} -\frac{1}{2}\ln \frac{\mu^2}{Q^2 t^2} -\frac{3}{2}+S_2 (M^2)\Bigr]
+\Bigl[ \frac{1}{2p^2} \sqrt{1-\frac{4p^2}{Q^2 t^2}}\Bigr]_{M^2}\Bigr\},
\end{equation}
where 
\begin{equation}
S_2 (M^2) = \sqrt{1- \frac{4M^2}{Q^2 t^2}} -\frac{1}{2}\ln \frac{w_+}{w_-},
\end{equation}
with $S_2 (Q^2 t^2/4) =0$.

The final collinear matrix element for the unintegrated jet function in the SW algorithm is written as
\begin{eqnarray}
M_n^{\mathrm{unint}} &=& \frac{\alpha_s C_F}{2\pi}  \Bigl\{ \delta(p^2) \Bigl[ \frac{1}{\euv^2} +\frac{1}{\euv} \Bigl( \frac{3}{2} 
+\ln \frac{\mu^2}{Q^2 t^2}\Bigr) +\frac{3}{2} \ln \frac{\mu^2}{Q^2 t^2}  +\frac{1}{2} \ln^2 \frac{\mu^2}{Q^2 t^2} \nonumber \\
&&+\frac{13}{2} -\frac{3}{4}\pi^2 + S(M^2)\Bigr] \nonumber \\
&+& \Bigl[ \frac{1}{p^2} \Bigl( -\frac{3}{2} \sqrt{1-\frac{4p^2}{Q^2 t^2}} +2 \ln \frac{1+\sqrt{1-4p^2/Q^2 t^2}}{1-\sqrt{1-4p^2/Q^2 t^2}} 
\Bigr)\Bigr]_{M^2} \Bigr\},
\end{eqnarray}
where $S(M^2)$ is the function which vanishes when $M=Qt/2$, and is given by
\begin{eqnarray}
S(M^2) &=& 2S_1 (M^2) + S_2 (M^2)  \nonumber \\
&=& 2 \mathrm{Li}_2 (w_+) -2 \mathrm{Li}_2 (w_-) -3\sqrt{1-4M^2/Q^2 t^2} +\frac{3}{2} \ln \frac{w_+}{w_-}  \nonumber \\
&+& \ln^2 w_- -\ln^2 w_+ -2 \ln \frac{M^2}{Q^2 t^2} \ln \frac{M^2/Q^2 t^2}{w_+ - M^2/Q^2 t^2}.  
\end{eqnarray}
The unintegrated jet function at order $\alpha_s$ in the SW jet algorithm is given by
\begin{eqnarray}
J_{\mathrm{SW}}^{(1)}  (p^2, Q,t,\mu)&=& \frac{\alpha_s C_F}{2\pi}  \Bigl\{ \delta(p^2) \Bigl(  
\frac{3}{2} \ln \frac{\mu^2}{Q^2 t^2}  +\frac{1}{2} \ln^2 \frac{\mu^2}{Q^2 t^2} +\frac{13}{2} -\frac{3}{4}\pi^2 \Bigr) \nonumber \\
&+& \Bigl[ \frac{1}{p^2} \Bigl( -\frac{3}{2} \sqrt{1-\frac{4p^2}{Q^2 t^2}} +2 \ln \frac{1+\sqrt{1-4p^2/Q^2 t^2}}{1-\sqrt{1-4p^2/Q^2 t^2}} 
\Bigr)\Bigr]_{M=Qt/2} \Bigr\},
\end{eqnarray}
Here the relation between the unintegrated and the integrated jet functions remains the same as in the cone-type algorithm.

\section{Soft function}
\begin{figure}[b] 
\begin{center}
\includegraphics[width=16cm]{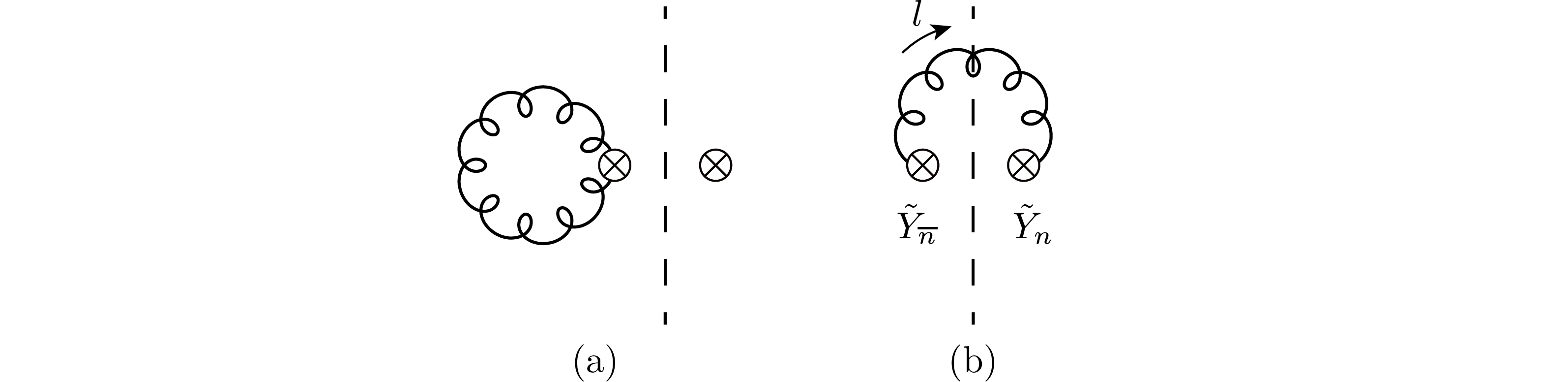}
\end{center}  
\vspace{-0.3cm}
\caption{Feynman diagrams for soft functions at one loop  
(a) virtual corrections and (b) real gluon emission.  The diagrams with the hermitian conjugate are omitted. \label{softfig}}
\end{figure}

The soft function with a jet algorithm is defined as
\begin{equation}
\mathcal{S}_{\Theta} =\sum_{X_s} \frac{1}{N_c} \mathrm{Tr} \langle 0| \tilde{Y}_{\overline{n}}^{\dagger}
\tilde{Y}_n \Theta_{\mathrm{soft}} \tilde{Y}_n^{\dagger} \tilde{Y}_{\overline{n}} |0\rangle.
\end{equation}
The Feynman diagrams for the soft function at order $\alpha_s$ are shown in Fig.~\ref{softfig}.  Fig.~\ref{softfig} (a) and (b) describe the virtual
and real contributions respectively and the dashed line represents the cut. The soft function is twice the contributions of Fig.~\ref{softfig} since the
hermitian conjugate should be added.

The virtual correction is given as
\begin{equation} \label{softv}
S_a = -\frac{\alpha_s C_F}{2\pi} (e^{\gamma_{\mathrm{E}}} \mu^2)^{\eps} \Gamma(1+\eps) \int_0^{\infty} du dv (uv)^{-1-\eps}
= - \frac{\alpha_s C_F}{2\pi} \Bigl( \frac{1}{\euv} -\frac{1}{\eir}\Bigr)^2.
\end{equation}

The real gluon emission can be computed according to the jet algorithm in Eq.~(\ref{jetsoft}). The phase space available for the soft function
is given in Fig.~\ref{softps}. First, the contribution in which the soft gluon is in the
jet in the $n$ direction is given by
\begin{equation}
M_1 =\frac{\alpha_s C_F}{2\pi} \frac{(e^{\gamma_{\mathrm{E}}} \mu^2)^{\eps}}{\Gamma(1-\eps)} \int_{2\beta Q}^{\infty} dl_- l_-^{-1-\eps} 
\int_0^{t^2 l_-} dl_+ l_+^{-1-\eps}. 
\end{equation}
Here the region $l_- <2 \beta Q$ is excluded since it will be included when we consider the case when the energy of the soft gluon is less than $\beta Q$.
Since the $l_-$ integral has the UV singularity, $\eps >0$,   the $l_+$ integral is rewritten as
\begin{equation}
\int_0^{t^2 l_-} dl_+ l_+^{-1-\eps} = \int_0^{\infty} dl_+ l_+^{-1-\eps} -\int_{t^2 l}^{\infty} l_+^{-1-\eps} =\frac{1}{\euv} -\frac{1}{\eir}
-\frac{1}{\euv} (t^2 l_-)^{-\eps}.
\end{equation}
Then $M_1$ becomes
\begin{equation}
M_1 = \frac{\alpha_s C_F}{2\pi} \Bigl[ \frac{1}{2\euv^2} -\frac{1}{\euv\eir}+\frac{1}{\euv} \ln t -\frac{1}{\eir} \ln \frac{\mu}{2\beta Q} 
- \ln^2 \frac{\mu}{2\beta Qt} +\frac{\pi^2}{24}\Bigr].
\end{equation}
The amplitude for the case in which the soft gluon in the $\overline{n}$ jet is the same as $M_1$.

The contribution from the soft gluon with energy less than $\beta Q$ is given as
\begin{eqnarray}
M_2 &=& \frac{\alpha_s C_F}{2\pi} \frac{(e^{\gamma_{\mathrm{E}}} \mu^2)^{\eps}}{\Gamma(1-\eps)} \int_0^{2\beta Q} dl_+ l_+^{-1-\eps}
\int_0^{2\beta Q-l_+} dl_- l_-^{-1-\eps} \nonumber \\
&=& \frac{\alpha_s C_F}{2\pi} \Bigl( \frac{1}{\eir^2} +\frac{2}{\eir} \ln \frac{\mu}{2\beta Q} + 2 \ln^2 \frac{\mu}{2\beta Q} -\frac{\pi^2}{4} 
\Bigr).
\end{eqnarray}
Summing over all the contributions, the real contribution is given by
\begin{equation}
S_b = 2M_1 +M_2 =\frac{\alpha_s C_F}{2\pi}\Bigl[ \Bigl(\frac{1}{\euv} -\frac{1}{\eir}\Bigr)^2 +\frac{2}{\euv} \ln t +2 \ln^2 \frac{\mu}{2\beta Q}
- 2\ln^2 \frac{\mu}{2\beta Qt} -\frac{\pi^2}{6}\Bigr].
\end{equation}
There are two small regions missing in the above computation $2\beta Q/(1+t^2) < l_-, l_+<2 \beta Q$ in Fig.~\ref{softps}, but 
the contributions from those regions are negligible for small $t\sim \mathcal{O} (\lambda^2)$.

The amplitude for the soft function is given by
\begin{equation}
M_{\mathrm{soft}} = 2(S_a +S_b) = \frac{\alpha_s C_F}{\pi} \Bigl( \frac{2}{\euv} \ln t + 4 \ln \frac{\mu}{2\beta Q} \ln t
-2 \ln^2 t -\frac{\pi^2}{6} \Bigr),
\end{equation}
and the integrated soft function at order $\alpha_s$ is given as
\begin{equation}
\mathcal{S}_{\Theta}^{(1)} (t, \beta,\mu) =   \frac{\alpha_s C_F}{\pi} \Bigl(  4 \ln \frac{\mu}{2\beta Q} \ln t -2 \ln^2 t -\frac{\pi^2}{6}
\Bigr).
\end{equation}

\section{Jet cross sections}
Let us summarize all the results for the factorized parts before we discuss the dijet cross sections. The hard function at order $\alpha_s$
is  given from Eq.~(\ref{hard}) as
\begin{equation} 
H^{(1)}(Q^2,\mu) = \frac{\alpha_s C_F}{2\pi}
\Bigl (-\ln^2  \frac{\mu^2}{Q^2} -3 \ln \frac{\mu^2}{Q^2} -8
+\frac{7\pi^2}{6}\Bigr). 
\end{equation}
The integrated jet functions  at order $\alpha_s$ are given by
\begin{eqnarray}
\mathcal{J}_{\mathrm{cone}}^{(1)}(Q, t,\mu) &=&\frac{\alpha_s C_F}{2\pi} \Bigl( \frac{3}{2} \ln \frac{\mu^2}{Q^2 t^2} 
+\frac{1}{2} \ln^2 \frac{\mu^2}{Q^2 t^2} +\frac{7}{2} +3\ln 2 -\frac{5\pi^2}{12}\Bigr),  \nonumber \\
\mathcal{J}_{\mathrm{SW}}^{(1)} (Q, t,\mu) &=& \frac{\alpha_s C_F}{2\pi} \Bigl(
\frac{3}{2} \ln \frac{\mu^2}{Q^2 t^2} +\frac{1}{2} \ln^2 \frac{\mu^2}{Q^2 t^2} +\frac{13}{2} -\frac{3}{4} \pi^2\Bigr). 
\end{eqnarray}
The soft function at order $\alpha_s$ is given as
\begin{equation}
\mathcal{S}^{(1)}_{\Theta} (t,\beta,\mu) = \frac{\alpha_s C_F}{\pi} \Bigl(   4 \ln \frac{\mu}{2\beta Q} \ln t -2 \ln^2 t -\frac{\pi^2}{6} \Bigr).
\end{equation}

The anomalous dimensions of the hard, collinear and soft functions are given as
\begin{eqnarray}
\gamma_H &=& \frac{\alpha_s C_F}{2\pi} \Bigl( -8 \ln \frac{\mu}{Q}-6\Bigr), \nonumber \\
\gamma_J &=& \frac{\alpha_s C_F}{2\pi} \Bigl( 4 \ln \frac{\mu}{Qt} +3\Bigr), \nonumber \\
\gamma_S &=& \frac{\alpha_s C_F}{2\pi} 8 \ln t,
\end{eqnarray}
which yields
\begin{equation}
\gamma_H + 2\gamma_J +\gamma_S =0.
\end{equation}
It means that the dijet cross section is independent of the renormalization scale $\mu$ to
next-to-leading log order. Explicitly at order $\alpha_s$, the dijet cross section in the SW algorithm  is given by \cite{Sterman:1977wj}
\begin{eqnarray}
\sigma_{\mathrm{SW}}^{(1)} &=& \sigma_0 \Bigl(H^{(1)} (Q^2,\mu) +2 \mathcal{J}_{\mathrm{SW}}^{(1)} (Q, t, \mu)
+ \mathcal{S}^{(1)} (t, \beta, \mu)\Bigr)\nonumber \\
&=& \sigma_0  \frac{\alpha_s C_F}{\pi} \Bigl( -4 \ln 2 \beta \ln t -3\ln t +\frac{5}{2} -\frac{\pi^2}{3}\Bigr), 
\end{eqnarray}
 whereas the dijet cross section in the cone-type algorithm differs only in the constant term which is $-1/2+3\ln 2$ instead of $5/2-\pi^2/3$. 

\section{Discussion on various jet functions}
There have been many forms  and definitions about the jet functions. Here we have confirmed 
that the integrated jet function is obtained by integrating the unintegrated
jet function. On the other hand, the unmeasured jet function in Ref.~\cite{Ellis:2010rwa} is not obtained by integrating the measured jet function. 
Especially, the UV divergence, or the corresponding anomalous dimensions are different in the measured and unmeasured jet functions.
This may be confusing at first sight, but the definitions and the power counting are different, therefore there is no inconsistency. For example, when
the total dijet cross section is computed, the calculations using the unintegrated, integrated jet functions, and the measured, unmeasured jet functions
(with the corresponding soft functions)
yield the same result. Here we discuss the origins of the difference and physics in detail. 

First, the inclusive jet function is the jet function without any jet algorithm and it has been computed to
one loop \cite{Bauer:2003pi,Bosch:2004th,Chay:2013zya}, and to two loops \cite{Becher:2006qw}. At one loop, the divergent part of the inclusive 
 jet function is given as
\begin{equation} \label{injet}
J^{(1)}_{\mathrm{incl}} (\mu) = \frac{\alpha_s C_F}{2\pi} \delta (p^2)\Bigl[ \frac{2}{\euv^2} +\frac{1}{\euv} \Bigl(\frac{3}{2} +2\ln \frac{\mu^2}{Q^2} \Bigr)
 \Bigr],
\end{equation}
where only the divergent terms proportional to $\delta (p^2)$ are shown. In this form, it is easy to compare the difference of the 
anomalous dimensions when integrated over the relevant variables.

The corresponding divergent terms in the unintegrated jet function and the measured and unmeasured  jet functions 
 with the cone-type algorithm are given as
\begin{eqnarray}\label{unjet}
J ^{(1)}  (p^2,\mu) &=&\frac{\alpha_s C_F}{2\pi}\delta(p^2) \Bigl[  \frac{1}{\euv^2} +\frac{1}{\euv} \Bigl( \frac{3}{2}  
+\ln \frac{\mu^2}{Q^2 t^2}\Bigr)  \Bigr],  \nonumber \\
J^{(1)}_{\mathrm{unmeas}} (\mu^2) &=&\frac{\alpha_s C_F}{2\pi} \Bigl[\frac{1}{\euv^2} +\frac{1}{\euv} \Bigl(
\frac{3}{2} +\ln \frac{\mu^2}{Q^2 t^2}\Bigr) \Bigr], \nonumber \\
J^{(1)}_{\mathrm{meas}} (\tau_0,\mu^2) &=&   \frac{\alpha_s C_F}{2\pi}\delta(\tau_0) \Bigl[ \frac{2}{\euv^2} +\frac{1}{\euv} \Bigl(
\frac{3}{2} +2\ln \frac{\mu^2}{Q^2} \Bigr) \Bigr],
\end{eqnarray}
where $\tau_0 =p^2/Q^2$ for $a=0$ in Ref.~\cite{Ellis:2010rwa}.
In comparing the differences, it is convenient to note the coefficients of the double poles of the various jet functions.

\begin{figure}[b] 
\begin{center}
\includegraphics[width=11cm]{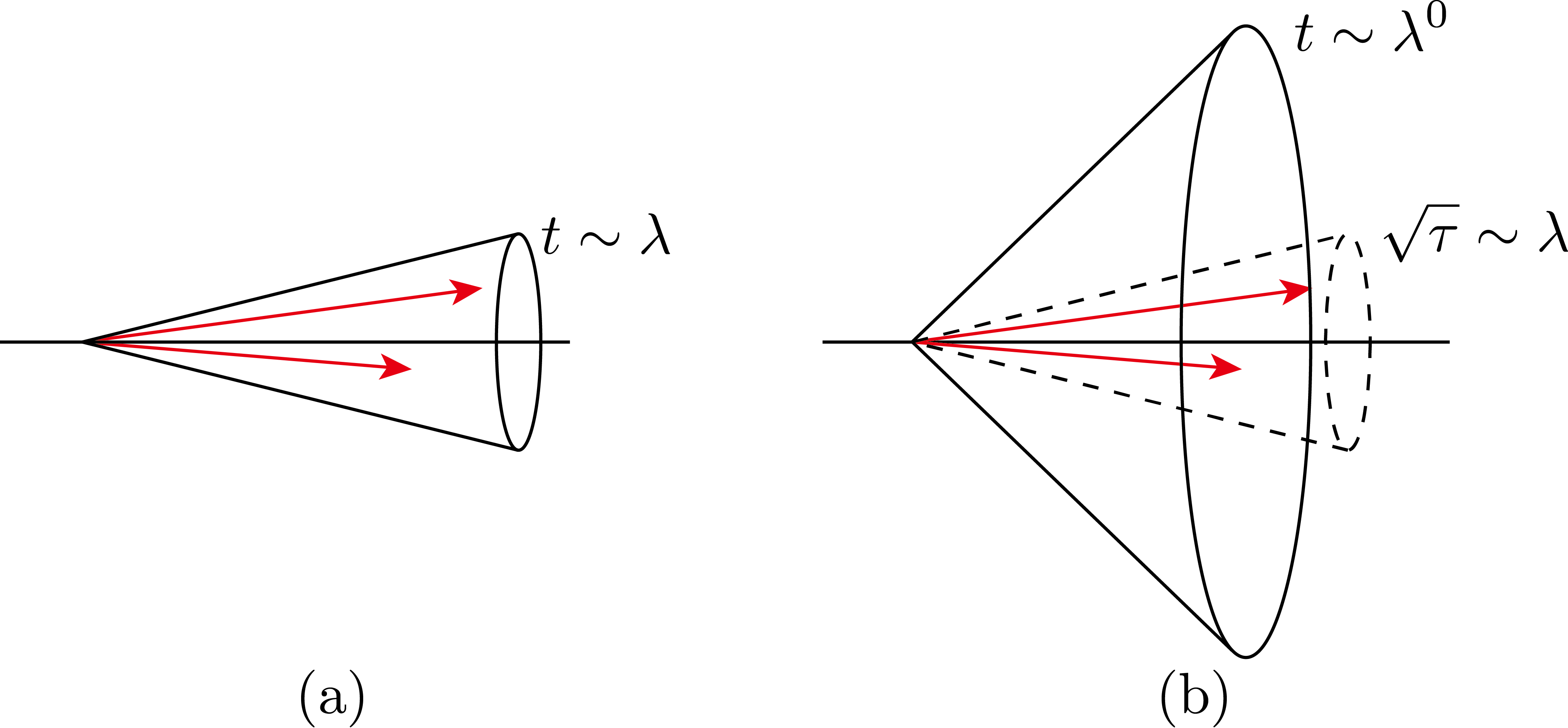}
\end{center}  
\vspace{-0.3cm}
\caption{Comparison of the jet sizes according to the power counting of the parameters. (a) In the unintegrated jet function, the cone size 
(solid line) is of order
$t\sim \lambda$. (b) In the measured jet function, the jet cone size is of order $R\sim \lambda^0$, while the physical cone size (dashed line) is
 controlled by  $\sqrt{\tau} \sim \lambda$, which is of the same order as $t$ in (a).
\label{cones}}
\end{figure}

 The integrated jet function and the unmeasured jet function are the same including the finite terms. Actually these definitions are the same.
The difference lies in the unintegrated and the measured jet functions. In computing the unintegrated jet function, the jet algorithm requires that 
two particles be in the cone size of order $t\sim\lambda$. In contrast, the measured
jet function requires that the particles contributing to $\tau$ should be confined in a cone with the size of order $\sqrt{\tau}_a \sim \lambda$, 
while the cone size is approximately $R\sim \lambda^0$. It means that in the measured jet function, there is effectively no jet algorithm involved,
and the jet size is controlled only by the physical observable $\tau_a$. The difference of the jet configurations in the unintegrated jet function
 and in the measured jet function is sketched in Fig.~\ref{cones}. Therefore the idea of the inclusive jet function (with no jet algorithm) is employed
in the measured jet function as far as the jet algorithm is concerned. When compared, it is clear from Eqs.~(\ref{injet}) and (\ref{unjet}) 
that the divergent parts of the inclusive jet function and the measured jet function are the same.
This difference causes the different behavior of the divergences between the unintegrated and the measured jet functions.

Then there also arises the question of the behavior of the dijet cross section,which is a convolution of the hard coefficient (the same), the jet functions,
and the soft function. When the integrated or the unintegrated jet function is used, since their anomalous dimensions are the same, the dijet
cross section is the same, and independent of the renormalization scale $\mu$. However, it seems that the cross sections with the unmeasured jet 
function and the measured jet function integrated over $\tau$ may have different anomalous dimensions, which is troublesome. The answer lies
in the fact that the measured soft function is also affected by the jet size of $\sqrt{\tau} \sim \lambda$ with $t\sim\lambda^0$. 
If the change of the measured soft function is 
included, we have verified that the same dijet cross section is obtained. Therefore the dijet cross section remains the same
whichever jet functions with their corresponding soft functions are used.

\section{Conclusion}
The implementation of the jet algorithms in SCET successfully factorizes the dijet cross section in $e^+ e^-$ annihilation. We have confirmed that the 
divergences in each factorized part are truly of the UV origin in the cone-type and the SW jet algorithms. Thus the jet function and the soft function are
infrared finite in each jet algorithm. This is verified in pure dimensional regularization and in the regularization with rapidity regulator in the 
collinear Wilson line. The question still remains whether it is also true in various other jet algorithms such as 
the exclusive $k_T$ \cite{Catani:1991hj} and Georgi \cite{Georgi:2014zwa} algorithms. It  will be presented soon elsewhere \cite{kimchaychul}.

In proving that the jet and the soft functions are IR finite in pure dimensional regularization with the spacetime dimension $D=4-2\eps$, 
the IR and UV divergences are entangled due to the phase space constraint. The phase space constraints for the collinear, zero-bin and 
soft contributions complicate the identification of the sources of the divergence. 
We have performed a careful analysis to identify the source of the
divergences and to disentangle the IR and UV divergences. And in the cone-type and in the SW jet algorithms, the jet function and the soft function
are proven to be IR finite to next-to-leading order, which enables the factorization of the dijet cross section.

The integrated jet function is indeed obtained by integrating the unintegrated jet function. In this case, as was
discussed in comparing with the relation between the measured and unmeasured jet functions, the power counting of the parameters involved is 
important. The power counting for the (un)integrated jet function is such that $t\sim \lambda$, which is related to the cone size, and $\beta \sim 
\lambda^2$, which is related to the jet veto.

In calculating the unintegrated jet function, it is important to keep the power counting correctly. Especially in computing the zero-bin contribution
$M_b^0$, the distribution function should be concentrated near $p^2=0$ by power counting. 
This distinction makes the relation between the unintegrated
and the integrated jet functions correct. If we do not care about the power counting and compute the zero-bin contribution with the distribution
away from $p^2=0$, $p^2$ can have any value and the result becomes exactly the inclusive jet function with $t=1$, instead of the unintegrated jet function.  This is obvious because the inclusive jet function includes all the regions of $p^2$. However, if we employ the inclusive jet function, or
if we neglect the careful power counting, the dijet cross section would depend on the renormalization scale $\mu$, which is not sensible. 
Therefore careful power counting should be performed in computing 
jet functions.

It will be interesting to probe the divergence structure in other jet algorithms not only in $e^+ e^-$ annihilation but also in hadronic collisions. 
The kinematics in both cases is different, and the color structure is more complicated since the initial states also consist of hadrons. The factorization
property with various jet algorithms in different scattering processes will be 
pursued based on the analysis in this paper.

\appendix*
\section{Integrated jet functions in the SW jet algorithm with rapidity regulator}
Jet functions can be computed with a rapidity regulator for the IR divergence, while the UV divergence is handled by the dimensional regularization. For the integrated jet function, we use Eq.~(\ref{irreg}) in the
collinear Wilson line to regulate the IR divergence. The virtual corrections in Fig.~\ref{intjet} (a) depend on the 
regulator $\Delta$, but the net collinear contribution after the zero-bin subtraction yields the same result
as in Eq.~(\ref{cola}). 

In the SW jet algorithm, the naive collinear contribution in Fig.~\ref{intjet} (b) is given as
\begin{eqnarray}
\tilde{M}_b &=& \frac{\alpha_s C_F}{2\pi} \frac{(e^{\gamma_{\mathrm{E}}} \mu^2 )^{\eps}}{\Gamma (1-\eps)} \int dl_+ dl_- 
\frac{Q-l_-}{Q(l_- -\Delta)} l_-^{-\eps}  l_+^{-1-\eps} \Theta_J \nonumber \\
&=& \frac{\alpha_s C_F}{2\pi} \frac{(e^{\gamma_{\mathrm{E}}} \mu^2 )^{\eps}}{\Gamma (1-\eps)} \int_0^Q dl_- 
\frac{Q-l_- }{Q(l_- -\Delta)} l_-^{-\eps} \int_0^{l_- t^2 (Q-l_-)^2/Q^2} dl_+ l_+^{-1-\eps}\nonumber \\
&=& \frac{\alpha_s C_F}{2\pi}  \Bigl[  \Bigl(\frac{1}{\eir} + \ln \frac{\mu^2}{Q^2 t^2} \Bigr) \Bigl(1+ \ln \frac{-\Delta}{Q} \Bigr) - \ln^2 \frac{-\Delta}{Q}    + 4-\frac{2}{3}\pi^2 \Bigr].
\end{eqnarray}
For the zero-bin contribution, since the IR divergence is handled by the IR regulator, the $\eps$ should be 
of the UV origin, $\eps >0$. Therefore it yields
\begin{eqnarray}
M_b^0  &=& \frac{\alpha_s C_F}{2\pi} \frac{(e^{\gamma_{\mathrm{E}}} \mu^2 )^{\eps}}{\Gamma (1-\eps)} \int_0^{\infty} dl_- 
\frac{ l_-^{-\eps}}{l_- -\Delta}\int_0^{t^2l_- } dl_+ l_+^{-1-\eps}\nonumber \\
&=& \frac{\alpha_s C_F}{2\pi} \frac{(e^{\gamma_{\mathrm{E}}} \mu^2 )^{\eps}}{\Gamma (1-\eps)} \int_0^{\infty} dl_-  \frac{ l_-^{-\eps}}{l_- -\Delta} \Bigl(\int_0^{\infty } dl_+ l_+^{-1-\eps}
-\int_{t^2l_- }^{\infty}  dl_+ l_+^{-1-\eps} \Bigr)  \nonumber \\
&=&  \frac{\alpha_s C_F}{2\pi}
\Bigl( -\frac{1}{\euv\eir} +\frac{1}{2\euv^2} +\frac{1}{\euv} \ln t 
-\frac{1}{\eir} \ln \frac{\mu}{-\Delta} +2\ln \frac{\mu}{-\Delta} -\ln^2 \frac{\mu}{-\Delta} \nonumber \\
& -&\ln^2 t -\frac{7}{24}\pi^2\Bigr).
\end{eqnarray} 
And the net collinear contribution is given by
\begin{eqnarray}
M_b &=& \tilde{M}_b -M_b^0 \nonumber \\
&=& \frac{\alpha_s C_F}{2\pi}  \Bigl[ \frac{1}{\euv \eir} -\frac{1}{2\euv^2} -\frac{1}{\euv} \ln t 
+\frac{1}{\eir} \Bigl( 1+\ln \frac{\mu}{Q}\Bigr) +2\ln \frac{\mu}{Qt} +\ln^2  \frac{\mu}{Qt} \nonumber \\
&&+ 4-\frac{3}{8}\pi^2\Bigr].
\end{eqnarray}
Likewise, $M_c$ is given by
\begin{equation}
M_c = \frac{\alpha_s C_F}{2\pi}  \Bigl( -\frac{1}{\eir} -3-2\ln \frac{\mu}{Qt}\Bigr). 
\end{equation}
Finally the total collinear amplitude is given as
\begin{eqnarray} 
M_{\mathrm{coll}} &=& 2 (M_a +M_b) + M_c + Z_{\xi} +R_{\xi} \\
&=& \frac{\alpha_s C_F}{2\pi} \Bigl[ \frac{1}{\euv^2} +\frac{1}{\euv} \Bigl( \frac{3}{2} + \ln \frac{\mu^2}{Q^2 t^2} \Bigr)
+\frac{3}{2} \ln \frac{\mu^2}{Q^2 t^2} +\frac{1}{2} \ln^2 \frac{\mu^2}{Q^2 t^2} +\frac{13}{2} -\frac{3}{4} \pi^2\Bigr], \nonumber
\end{eqnarray}
which is the same as Eq.~(\ref{swjet}).

\acknowledgments
J.~Chay and I.~Kim are grateful to Christopher Lee for the discussion to clarify the differences between
the integrated, unintegrated jet functions and the unmeasured, measured jet functions while the authors
visited Los Alamos National Laboratory.
J. Chay and I. Kim are supported by Basic Science Research Program through the National Research Foundation of Korea(NRF) funded 
by the Ministry of Education(NRF-2012R1A1A2008983, NRF-2014R1A1A2058142).  
C.~Kim is supported by Basic Science Research 
Program through the National Research Foundation of Korea(NRF) funded by the Ministry of Science, ICT and Future Planning(NRF-2012R1A1A1003015, NRF-2014R1A2A1A11052687).

\end{document}